\documentclass[twocolumn,showpacs,prl,superscriptaddress,nofootinbib]{revtex4-1}
\usepackage{graphicx}
\usepackage{amsmath}

\newcommand{\bra}[1]{\mbox{$\left\langle #1 \right|$}}
\newcommand{\ket}[1]{\mbox{$\left| #1 \right\rangle$}}

\begin{document}

\title{Long-Distance Measurement-Device-Independent Multiparty Quantum
Communication}
\author{Yao Fu}
\affiliation{Hefei National Laboratory for Physical Sciences at Microscale and Department
of Modern Physics, University of Science and Technology of China, Hefei,
Anhui 230026, People's Republic of China}
\affiliation{The CAS Center for Excellence in QIQP and the Synergetic Innovation Center
for QIQP, University of Science and Technology of China, Hefei, Anhui
230026, People's Republic of China}
\author{Hua-Lei Yin}
\affiliation{Hefei National Laboratory for Physical Sciences at Microscale and Department
of Modern Physics, University of Science and Technology of China, Hefei,
Anhui 230026, People's Republic of China}
\affiliation{The CAS Center for Excellence in QIQP and the Synergetic Innovation Center
for QIQP, University of Science and Technology of China, Hefei, Anhui
230026, People's Republic of China}
\author{Teng-Yun Chen}
\affiliation{Hefei National Laboratory for Physical Sciences at Microscale and Department
of Modern Physics, University of Science and Technology of China, Hefei,
Anhui 230026, People's Republic of China}
\affiliation{The CAS Center for Excellence in QIQP and the Synergetic Innovation Center
for QIQP, University of Science and Technology of China, Hefei, Anhui
230026, People's Republic of China}
\date{\today}
\author{Zeng-Bing Chen}
\email{zbchen@ustc.edu.cn}
\affiliation{Hefei National Laboratory for Physical Sciences at Microscale and Department
of Modern Physics, University of Science and Technology of China, Hefei,
Anhui 230026, People's Republic of China}
\affiliation{The CAS Center for Excellence in QIQP and the Synergetic Innovation Center
for QIQP, University of Science and Technology of China, Hefei, Anhui
230026, People's Republic of China}
\date{\today}

\begin{abstract}
The Greenberger-Horne-Zeilinger (GHZ) entanglement, originally introduced to
uncover the extreme violation of local realism against quantum mechanics, is
an important resource for multiparty quantum communication tasks. But the
low intensity and fragility of the GHZ entanglement source in current
conditions have made the practical applications of these multiparty tasks an
experimental challenge. Here we propose a feasible scheme for
practically distributing the post-selected GHZ entanglement over a distance
of more than 100 km for experimentally accessible parameter regimes.
Combining the decoy-state and measurement-device-independent protocols for
quantum key distribution, we anticipate that our proposal suggests an
important avenue for practical multiparty quantum communication.
\end{abstract}

\maketitle



Remote distribution of quantum signals (photonic states) is an essential
task in the realm of quantum communication. Quantum key distribution (QKD)
allows the information-theoretically secure transmission of classical
messages and requires delivery of either single photons in the case of BB84
protocol \cite{BB_84}, or entangled photons in the case of Ekert91 protocol
\cite{ekert:1991:quantum}. Remote distribution of entanglement also enables
certain classically impossible tasks, such as quantum teleportation of
unknown states and quantum dense coding \cite{Pan:Multiphoton:2012}. Up to now, tremendous efforts have been
dedicated to increase the transmission distance of quantum
communication between \textit{two} legitimate users. The recorded distance
for QKD has been more than 300 km for standard telecom
fiber links \cite{Shibata2014}, while quantum
teleportation has been demonstrated over a distance of more than 100 km for
free-space channels \cite{Yin:2012:quantum:Ma:2012:quantum}.

So far, most theoretical and experimental works on quantum communication are
focused on two-party protocols. Yet, multiparty quantum communication protocols do exist,
as illustrated by the fascinating examples like quantum
cryptographic conferencing (QCC) \cite%
{Bose:1998:Multiparticle,Chen:2007:Multi-pattite}, quantum secret sharing
(QSS) \cite{Hillery:1999:quantum,Cleve:QSS:1999,Tittel:2001:Exp:Chen:2005:QSS:Gaertner:2007:exp:Schmid:Exp:2005,Bell:2014:experimental} and
third-man quantum cryptography \cite{Zukowski:1998:Quest}. These multiparty
protocols require an important resource--the Greenberger-Horne-Zeilinger
(GHZ) entangled states \cite{GHZ:1989:GHZ,Mermin:1990:Inequality} with perfect multiparty quantum
correlations, which are originally introduced to reveal the extreme
violation of local realism against quantum mechanics. Nevertheless, the
practical applications of GHZ states are quite limited due to the lack of
two important factors--the high-intensity source and remote reliable
distribution of the GHZ states. The existing experimental works \cite%
{Tittel:2001:Exp:Chen:2005:QSS:Gaertner:2007:exp:Schmid:Exp:2005} on multiparty quantum communication remain
the proof-of-principle demonstration and reported rather low key rates. The
experimental distribution of the GHZ entanglement \cite{Erven:2014:EXP} was
achieved only recently, over a distance of less than 1 km for each
party of the GHZ-entangled photons. Thus, the current status of multiparty
quantum communication still remains an extreme experimental challenge even
under the state-of-the-art technologies and is far from practical
applications. In this Letter, we propose a feasible scheme for distributing
the post-selected GHZ entanglement over a distance of more than
100 km for experimentally relevant parameter regimes. Combining the
decoy-state QKD \cite{Hwang:2003:Quantum:Wang:2005:Beating:Lo:2005:Decoy}
and the measurement-device-independent (MDI) QKD \cite{Lo:MIQKD:2012:Braunstein:2012:Side}
technologies, our findings manifest the possibility for practical
applications of MDI multiparty quantum communication such as QCC and QSS,
as well as for the long-distance GHZ experiment.

\begin{figure*}[tbh]
\centering
\resizebox{13cm}{!}{\includegraphics{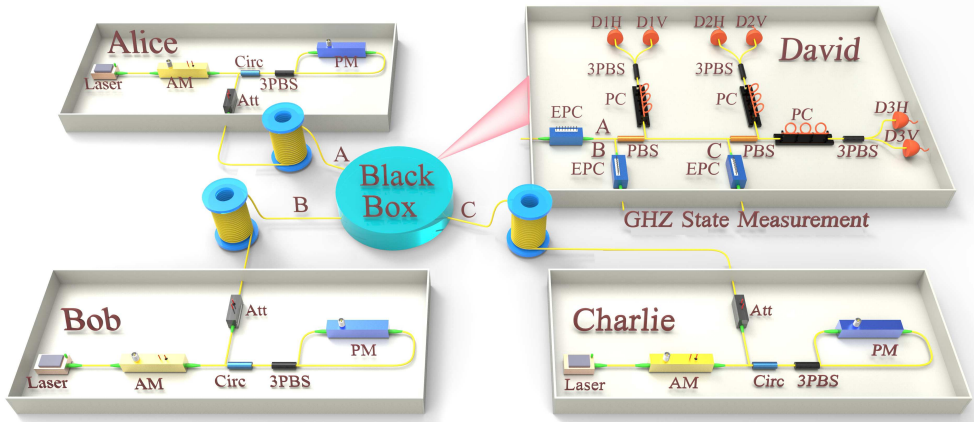}}
\caption{(color online) Schematic layout of the MDI-QCC setup. AM: amplitude modulator used to prepare
decoy states; 3PBS: 3-port
polarization-maintaining PBS, which, besides the function of PBS, can
transit optical pulses from fast axis to slow axis; Circ: circulator; PM: phase modulator, combining with 3PBS and Circ, is used to encode qubits; PC: polarization controller which
makes a unitary transformation like a half-wave plate such that it
corresponds to a $45^{\circ }$ rotation of the polarization; Black Box: the GHZ-state
measurement device; Att: attenuator used to prepare weak coherent pulses; EPC: electric polarization
controller used to adjust the frame of reference; PBS: polarizing
beam-splitter which transmits \mbox{$\left| H \right\rangle$} and reflects\mbox{$\left|V\right\rangle$} polarizations;  D1H,
D2V, D2H D2V, D3H and D3V: single-photon detectors.}
\label{Fig:Setup}
\end{figure*}

Multiparty quantum communication protocols aim to provide
information-theoretic security for highly sensitive and confidential multiuser communication based on the laws of quantum mechanics, which
physically outperform their classical counterparts. Their applications \cite{Hillery:1999:quantum,Cleve:QSS:1999,Bell:2014:experimental} range from the secret multiparty conference, remote voting, online auctioning, master key of the payment system, jointly checking
accounts containing quantum money \cite{Wiesner:1983}, to secure distributed quantum computation \cite{gottesman:1999:demonstrating}. Among them, QCC is a protocol for multiparty QKD \cite{Bose:1998:Multiparticle}, which requires a common random bit sequence (the keys)
to be securely shared among the legitimate users even in
the presence of any eavesdropper. QSS is a
protocol of splitting a message into several parts amongst a group of
participants, each of whom is allocated a share of the secret \cite%
{Hillery:1999:quantum}. As a consequence, only the entire set is sufficient
to read the message thoroughly. For example, QSS can be used to
guarantee that no single person can launch a nuclear missile, or open a bank
vault, but all legitimate users together can.

Before we describe our multiparty communication schemes in detail, let us
recapitulate the significance of the GHZ state $%
\mbox{$\left| \Phi_0^{\pm}
\right\rangle$}=1/\sqrt{2}(\mbox{$\left| HHH \right\rangle$}\pm
\mbox{$\left|
VVV \right\rangle$})$, where $\mbox{$\left| H \right\rangle$}$ and $%
\mbox{$\left| V \right\rangle$}$ represent photonic horizontal and vertical
polarizations, respectively. If three members of a GHZ state are measured
along $Z$ basis, each of them will give a random outcome, $%
Z_{A},~Z_{B},~Z_{C}$, and the outcomes of the three members will always be in
perfect correlations, $Z_{A}=Z_{B}=Z_{C}$, which can be used for multiparty
quantum cryptographic conferencing. Likewise, when three members of a GHZ state $%
\mbox{$\left| \Phi_{0}^{+} \right\rangle$}$ ($%
\mbox{$\left| \Phi_{0}^{-}
\right\rangle$}$) are measured along $X$ basis, each will give a random
outcome i.e., $X_{A},~X_{B},~X_{C}$, whose sharing of a binary correlation $%
X_{A}=X_{B}\oplus X_{C}$ ($X_{A}\oplus 1=X_{B}\oplus X_{C}$) will always
hold and can then be used for multiparty QSS. Besides, when Alice announces
her measurement result $X_{A}$, Bob and Charlie will have a perfect correlation
which can be used for third-man quantum cryptography.

Here we exploit an approach that requires neither the preparation in advance
nor the distribution of high-fidelity GHZ entangled states through a long
distance. The design is to take advantage of post-selected GHZ
states among three legitimate users (typically called Alice, Bob and
Charlie) to perform information-theoretically secure multiparty quantum
communication. Like the MDI-QKD protocol \cite%
{Lo:MIQKD:2012:Braunstein:2012:Side}, the post-selecting measurement device
here can be regarded as a black box which can be manipulated by anyone, even
the eavesdropper. Therefore, our scheme is naturally immune to all
detection-side attacks and can be regarded as the
combination of time-reversed GHZ state distribution and measurement.
Together with the decoy-state method \cite%
{Hwang:2003:Quantum:Wang:2005:Beating:Lo:2005:Decoy}, in which pulses with
different amplitudes are randomly mixed and phases are randomized, our scheme is able to
defeat photon-number-splitting attacks \cite{PNS:2000}. We utilize
conventional laser sources to obtain a long distribution distance between the middle node and
users for both the MDI-QCC and MDI-QSS protocols. Similarly
to the security proof of QKD \cite{Lo:MIQKD:2012:Braunstein:2012:Side,Lo:1999:unconditional:Shor:2000:Simple:Lo:2001:Sixstate}, we use multiparty
entanglement purification technique \cite{BDSW:1996:Mixed:Maneva:2000:Improved:Cirac:1999:se:1} to provide
information-theoretically secure information transmission. The security of our protocols is analyzed in the
Supplemental Material \cite{supplemental:material}.

In the following, let us explain our MDI-QCC and MDI-QSS protocols in more
details. The main quantum procedures of the two schemes are the same, while the
difference lies in their classical post-processing. The MDI-QCC (MDI-QSS) protocol uses the data in $Z$ ($X$) basis to extract secure keys.
Our setup is depicted in Fig.~\ref{Fig:Setup}. Here, we take MDI-QCC protocol as an
example. Alice, Bob and Charlie independently and randomly prepare quantum
states with phase-randomized weak coherent pulses in two complementary bases
($Z$ basis and $X$ basis). They send the pulses to the untrusted
fourth-party located in the middle node, David, to perform a GHZ-state
measurement which projects the incoming signals onto a GHZ state. Such a
measurement can be realized, for instance, using only linear optical
elements \cite{Pan:1999:GHZ}. Actually, this procedure only identifies two
of the eight GHZ states, while the identification of any one GHZ state is enough
to prove the security. A successful GHZ-state measurement corresponds to the
observation of three out of six detectors being clicked simultaneously. The
clicks in D1H, D2H and D3H, or in D1H, D2V and D3V, or in D1V, D2H and D3V,
or in D1V, D2V and D3H, imply a projection onto the GHZ state $%
\mbox{$\left|
\Phi_{0}^{+} \right\rangle$}=1/\sqrt{2}(\mbox{$\left| HHH \right\rangle$}+%
\mbox{$\left| VVV \right\rangle$})$, while the clicks in D1H, D2H and D3V,
or in D1H, D2V and D3H, or in D1V, D2H and D3H, or in D1V, D2V and D3V,
indicate a projection onto the GHZ state $%
\mbox{$\left| \Phi_{0}^{-}
\right\rangle$}=1/\sqrt{2}(\mbox{$\left| HHH \right\rangle$}-%
\mbox{$\left|
VVV \right\rangle$})$. David announces the events through public channels
whether he has obtained a GHZ state and which GHZ state he has received.
Alice, Bob and Charlie only keep the raw data of successful GHZ-state
measurements and discard the rest. They post-select the events where they use the
same basis in their transmission through an authenticated public channel.
Notice that Alice performs a bit flip when Alice, Bob and Charlie all choose
$X$ basis and David obtains a GHZ state $%
\mbox{$\left| \Phi_{0}^-
\right\rangle$}$. We employ the data of $Z$ basis to generate the
cryptographic conferencing keys, while the data of $X$ basis are totally used
to estimate errors. Alice, Bob and Charlie estimate the gain and quantum bit
error rate with decoy-state method, given that all of them send out
single-photon states. Afterwards, they extract secure cryptographic
conferencing keys after classical error correction and privacy amplification.

In the asymptotic limit, the
MDI-QCC key generation rate is given by \cite%
{Lo:MIQKD:2012:Braunstein:2012:Side,BDSW:1996:Mixed:Maneva:2000:Improved:Cirac:1999:se:1,GLLP:2004:Security}
\begin{equation}
\begin{aligned} \label{QCC:key:rate}
R_{QCC}=&Q_{v}^{Z}+Q_{111}^{Z}[1-H(e_{111}^{BX})]-H(E_{\mu\nu\omega}^{Z*})fQ_{\mu\nu%
\omega}^{Z}, \end{aligned}
\end{equation}%
where $Q_{\mu \nu \omega }^{Z}$ ($E_{\mu \nu \omega }^{Z\ast }$), the gain
(quantum bit error rate) of $Z$ basis, can be directly obtained from the
experimental results. The subscript $\mu \nu \omega $ means that Alice, Bob
and Charlie send out phase-randomized weak coherent pulses with intensity
$\mu $, $\nu $ and $\omega $, respectively. Note that each of these pulses
has single-photon state components and the ones of $n$ ($>1$) photons or zero photon.
For the post-selected GHZ states contributed solely by the single-photon
state components, the gain $Q_{111}^{Z}$ of $Z$ basis and the bit error rate
$e_{111}^{BX}$ of $X$ basis can be estimated by the decoy-state method. $Q_{v}^{Z}$ is the gain that Alice sends out vacuum state component in $Z$ basis and David obtains a GHZ state measurement result.
Here, we assume that Alice's raw key is the reference raw key, the parameter $f$ is the error correction efficiency ($f=1.16$ in our
simulation below), and $H(x)=-x\log _{2}x-(1-x)\log
_{2}(1-x)$ is the binary Shannon entropy function. The information-theoretic
security proof of MDI-QCC is shown in the Supplemental Material, from which
we have $E_{\mu \nu \omega }^{Z\ast }=\mathrm{max}$\{$E_{\mu \nu \omega
}^{ZAB}$,~$E_{\mu \nu \omega }^{ZAC}$\}. Here, $E_{\mu \nu \omega }^{ZAB}$
($E_{\mu \nu \omega }^{ZAC}$) is the quantum bit error rate of $Z$ basis
between Alice and Bob (Charlie).

In the same manner, the key generation rate of MDI-QSS in the asymptotic
limit is given by
\begin{equation}
\begin{aligned} \label{QSS:key:rate}
R_{QSS}=&Q_{v}^{X}+Q_{111}^{X}[1-H(e_{111}^{BZ})]-H(E_{\mu\nu\omega}^{X})fQ_{\mu\nu%
\omega}^{X}, \end{aligned}
\end{equation}%
where $Q_{\mu \nu \omega }^{X}$ ($E_{\mu \nu \omega }^{X}$), the gain
(quantum bit error rate) of $X$ basis, can also be directly obtained from
the experimental results. For the single-photon state contribution, the gain
$Q_{111}^{X}$ of $X$ basis and bit error rate $e_{111}^{BZ}$ of $Z$ basis
can be estimated by the decoy-state method. $Q_{v}^{X}$ is the gain that Alice sends out vacuum state component in $X$ basis and David obtains a GHZ state measurement result.
However, the overall quantum bit
error rate $E_{\mu \nu \omega }^{X}$ (always about $37.5\%$ for
arbitrarily-long transmission distances) in $X$ basis is so high that it is
virtually impossible to use weak coherent sources to perform MDI-QSS with Eq.~\eqref{QSS:key:rate}. To solve the problem, in the Supplemental Material we propose, in details, to use the triggered spontaneous parametric down conversion sources \cite{Lutkenhaus:2000:security}, or the conventional weak coherent state sources together with the quantum non-demolition measurement technique \cite{Grangier:QND:1998:mizutani:2014:measurement}.

However, such a solution is disadvantageous as it requires experimentally challenging technology. Fortunately, we can exploit the extra classical bit information \cite{ma:2012:improved,Ma:Alternative:2012} to extract the raw key with little bit error rate (almost zero) so that we can implement MDI-QSS, again with weak coherent sources.
The classical bit information corresponds to the information denoted by different overall phase regions over $[0,2\pi)$ (the phase post-selection technique).
Meanwhile, we assume the gain and bit error rate of single-photon states to be in a uniform distribution over $[0,2\pi)$ \cite{Ma:Alternative:2012}. Therefore, the secure key rate of MDI-QSS with phase post-selection can be given by (see Supplemental Material \cite{supplemental:material} for details)
\begin{equation}
\begin{aligned} \label{QSS:key:rate:2}
\widetilde{R}_{QSS}\geq &\frac{1}{K^2}Q_{111}^{X}[1-H(e_{111}^{BZ})]-H(\widetilde{E}_{\mu\nu\omega}^{X})f\widetilde{Q}_{\mu\nu\omega}^{X},
\end{aligned}
\end{equation}
where $K$ is the number of phase regions, $\widetilde{Q}_{\mu\nu\omega}^{X}$ and $\widetilde{E}_{\mu\nu\omega}^{X}$ are the gain and bit error rate of the pulses whose information is used to extract the raw key with little bit error rate. The phase post-selection technique requires to share a common phase reference \cite{Arrazola:2014:fingerprinting:Arrazola:2014:QDS:Dunjko:2014:QDS} among users. A method for distributing such a phase reference is suggested in Supplemental Material \cite{supplemental:material}. We note that the rigorous security of protocols involving phase post-selection technique needs more investigations in the contexts of both QKD \cite{ma:2012:improved,Ma:Alternative:2012} and MDI-QSS.

\begin{figure}[tbh]
\centering
\resizebox{7cm}{!}{\includegraphics{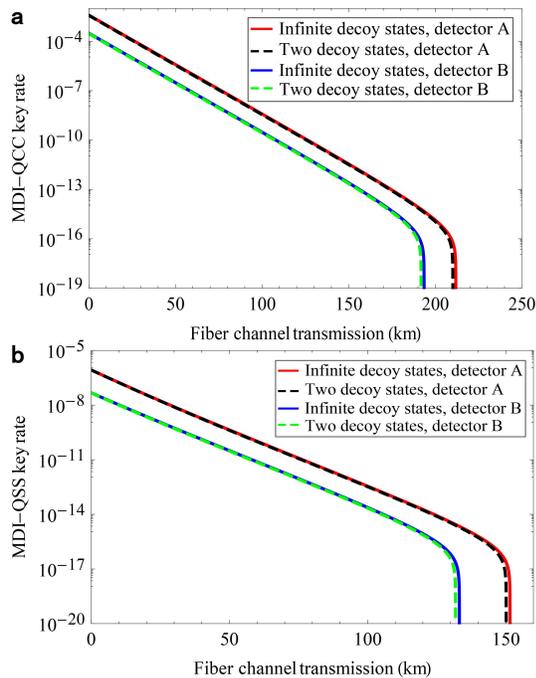}}
\caption{(color online) Lower bound on the secure key rates versus fiber
channel transmission. \textbf{a}, MDI-QCC with weak coherent sources.
\textbf{b}, MDI-QSS with weak coherent sources based on phase post-selection technique ($K=8$). We show the simulation results of infinite decoy states and two decoy states with detector A (B) of detection efficiency $93\%$ ($40\%$%
), respectively. The phase-randomized weak coherent sources with (without) the phase post-selection technique are used for MDI-QSS (MDI-QCC). The
intensity of the signal state and one decoy state is $0.4$ and $0.005$ ($0.11$ and $0.005$), while
the other decoy state is a vacuum state in MDI-QCC (MDI-QSS).}
\label{Fig:2}
\end{figure}

To analyze the performance of the secret key rates of MDI-QCC and MDI-QSS,
we present an analytical method with two decoy states to estimate the
relevant parameters $Q_{111}^{Z}$, $Q_{111}^{X}$, $e_{111}^{BZ}$ and $%
e_{111}^{BX}$, which are required to be evaluated in Eqs.~\eqref{QCC:key:rate}-%
\eqref{QSS:key:rate:2}. In our simulation, we
employ the following experimental parameters: the intrinsic loss
coefficient $\beta $ of the standard telecom fiber channel is $0.2$ dB/km.
For the threshold single-photon detectors, the detection efficiency $\eta
_{d} = 40\%$, and the background count rate $p_{d} = 1\times 10^{-7}$,
as used in a recent decoy-state MDI-QKD experiment \cite{Tang:MDI:2014}.
As a comparison, we also use the state-of-the-art single-photon detectors
\cite{Marsili:2013}, with $\eta _{d}=93\%$ and
$p_{d}=1\times 10^{-7}$. Here, we neglect the overall misalignment-error probability of the system. The secure key rates
of MDI-QCC with weak coherent sources in the cases of infinite decoy states
and of the two decoy states are shown in Fig.~\ref{Fig:2}a. From the simulation result, we see that the estimation
using two decoy states gives a secure key rate which is nearly the same as
the corresponding one using infinite decoy states. In the case of asymptotic data
with two decoy states, the secure transmission distance between Alice and
the middle node of MDI-QCC is about $190$ km for the detection efficiency of
$40\%$ ($210$ km for the detection efficiency of $93\%$). The secure key
rates of MDI-QSS with weak coherent sources based on overall phase post-selection technique are shown in Fig.~\ref%
{Fig:2}b. In the case of asymptotic data with two decoy states, the secure
transmission distance is about $130$ km for the detection efficiency of $%
40\%$ ($150$ km for the detection efficiency of $93\%$) between the middle
node and any user.

\begin{figure}[tbh]
\centering
\resizebox{7cm}{!}{\includegraphics{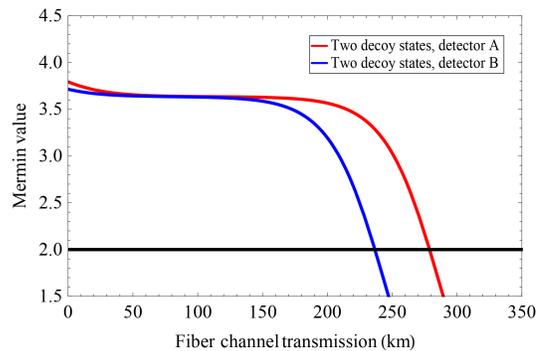}}
\caption{(color online) The Mermin value $M_{111}$ versus fiber channel
transmission. We use two decoy states to estimate $M_{111}$. We show
the simulation results for detector A (B) of detection efficiency of $93\%$
($40\%$) in red (blue) solid curve, respectively, the overall misalignment-error probability $e_{d}$ of the system is $1.5\%$, with other parameters identical to Fig.~\ref{Fig:2}a.
We also show the line of constant 2, which is the maximal value allowed by local realism.}
\label{Fig:3}
\end{figure}

The information-theoretic security of our multiparty quantum communication protocols is guaranteed by the GHZ entanglement purification technique \cite{BDSW:1996:Mixed:Maneva:2000:Improved:Cirac:1999:se:1} though the security of MDI-QSS is complicated by phase post-selection and needs further study. Indeed, the purpose of QCC and QSS protocols can be recognized as a procedure for Alice, Bob and Charlie to share almost perfect GHZ states. Qualitatively, the more perfect the GHZ entanglement shared by Alice, Bob and Charlie is, the more negligible the information would have been leaked to Eve \cite{Terhal:2004}. It is thus of vital importance to quantify the quality of the GHZ entanglement. For this purpose, Alice, Bob and Charlie independently and randomly prepare quantum
states with phase-randomized weak coherent pulses in two complementary bases
($X$ basis and $Y$ basis) and then send to David, who performs the GHZ-state
($\mbox{$\left| \Phi_{0}^{+} \right\rangle$}$) measurement.
What we take into consideration here is the post-selected GHZ states contributed
solely by the single-photon state components. This contribution can be estimated by the decoy-state method. For the GHZ entangled state $%
\mbox{$\left|
\Phi_{0}^{+} \right\rangle$}$, local realistic theories must obey Mermin's
inequality \cite{Mermin:1990:Inequality}:
\begin{equation}
\begin{aligned} \label{Merminvalue} M_{111}\equiv&{\langle XXX
\rangle}_{111}-{\langle XYY \rangle}_{111}\\ &-{\langle YXY
\rangle}_{111}-{\langle YYX \rangle}_{111}\leq2 . \end{aligned}
\end{equation}%
Here $M_{111}$ is defined as the Mermin value and witnesses the quality of the GHZ entanglement; ${\langle XXX\rangle }_{111}$
and so on are the expectation values with respect to the GHZ states solely contributed by the
single-photon state components. It is important to ensure that one only selects a single
ensemble corresponding to the successful projection
onto the GHZ state $\mbox{$\left| \Phi_{0}^{+} \right\rangle$}$.
In our post-selected GHZ states, the Mermin value, whose maximal value is 4 as predicted by quantum mechanics for ideal GHZ states,
can reach about $3.5$ as shown in
Fig.~\ref{Fig:3} over the distribution distance of about $170$ km from David to
Alice (Bob, Charlie); more details can be found in the Supplemental Material \cite{supplemental:material}. This indicates that high-quality GHZ entanglement can be generated at this distance by the protocol.
The proposed protocol can be regarded as a variance of the usual GHZ experiment testing local realism, namely, a time-reversed GHZ experiment where the state preparations replace the state measurements in the usual GHZ test. The interpretation of such a variance and, particularly, its relevance to the test of hidden-variable theories are interesting in its own right. We argue in the Supplemental Material \cite{supplemental:material} that such an experiment tests Mermin's argument \cite{Mermin:1990:Simple} on the Kochen-Specker theorem \cite{kochen:1967:problem}.

In summary, we propose a feasible protocol for distributing the
post-selected GHZ entanglement and MDI multiparty
quantum communication over a distance of more than 100 km for
experimentally accessible parameter regimes. Combining the decoy-state and MDI protocols
for QKD, we show that the
information-theoretically secure MDI-QCC with the conventional weak coherent
state sources can be implemented over a distance of about $190$ km, as well
as the MDI-QSS with weak coherent sources based on phase post-selection technique
over a distance of about $130$ km. These distances are significantly beyond what one could
expect previously for multiparty quantum communication with the GHZ
entanglement. Our proposal thus suggests an important avenue for practical
long-distance multiparty quantum communication. The extension of our scheme
to more legitimate users is straightforward.

We are grateful to the anonymous Referees for their valuable comments and suggestions
to improve the quality of the paper. This work has been supported by the CAS, the
NNSF of China under Grant No. 61125502, and
the Science Fund of Anhui Province for Outstanding Youth. Y.F.
and H.-L.Y. contributed equally to this work.


\bibliographystyle{apsrev}

\clearpage
\onecolumngrid

\section {Supplemental Material for ``Long Distance Measurement-Device-Independent Multiparty Quantum Communication"}

\section{I. Security analysis}\label{SM:EDP}
\subsection{A. GHZ State Entanglement Purification}

Here, the goal of an entanglement purification protocol is to distill nearly perfect GHZ states from noisy GHZ states initially shared among three distant parties (typically called Alice, Bob and Charlie). The density matrix $\rho_{ABC}$ describing Alice, Bob and Charlie's qubit system can be expressed in the GHZ basis \cite{Cirac:1999:se:1}, which is composed of eight orthogonal GHZ states:
\begin{equation}
\begin{aligned}  \label{GHZ:States:eq1}
\ket{\Phi_{0}^{+}}=\frac{1}{\sqrt{2}}(\ket{H}\ket{H}\ket{H}+\ket{V}\ket{V}\ket{V})=\frac{1}{2}(\ket{+++}+\ket{+--}+\ket{-+-}+\ket{--+}),\\
\ket{\Phi_{0}^{-}}=\frac{1}{\sqrt{2}}(\ket{H}\ket{H}\ket{H}-\ket{V}\ket{V}\ket{V})=\frac{1}{2}(\ket{++-}+\ket{+-+}+\ket{-++}+\ket{---}),\\
\ket{\Psi_{1}^{+}}=\frac{1}{\sqrt{2}}(\ket{V}\ket{H}\ket{H}+\ket{H}\ket{V}\ket{V})=\frac{1}{2}(\ket{+++}+\ket{+--}-\ket{-+-}-\ket{--+}),\\
\ket{\Psi_{1}^{-}}=\frac{1}{\sqrt{2}}(\ket{V}\ket{H}\ket{H}-\ket{H}\ket{V}\ket{V})=\frac{1}{2}(\ket{++-}+\ket{+-+}-\ket{-++}-\ket{---}),\\
\ket{\Psi_{2}^{+}}=\frac{1}{\sqrt{2}}(\ket{H}\ket{V}\ket{H}+\ket{V}\ket{H}\ket{V})=\frac{1}{2}(\ket{+++}-\ket{+--}+\ket{-+-}-\ket{--+}),\\
\ket{\Psi_{2}^{-}}=\frac{1}{\sqrt{2}}(\ket{H}\ket{V}\ket{H}-\ket{V}\ket{H}\ket{V})=\frac{1}{2}(\ket{++-}-\ket{+-+}+\ket{-++}-\ket{---}),\\
\ket{\Psi_{3}^{+}}=\frac{1}{\sqrt{2}}(\ket{H}\ket{H}\ket{V}+\ket{V}\ket{V}\ket{H})=\frac{1}{2}(\ket{+++}-\ket{+--}-\ket{-+-}+\ket{--+}),\\
\ket{\Psi_{3}^{-}}=\frac{1}{\sqrt{2}}(\ket{H}\ket{H}\ket{V}-\ket{V}\ket{V}\ket{H})=\frac{1}{2}(\ket{+-+}-\ket{++-}+\ket{-++}-\ket{---}).
\end{aligned}
\end{equation}
We take $\ket{\Phi_{0}^+}$ as the reference state in this paper. The GHZ state $\ket{\Phi_{0}^+}$ is stabilized by its stabilizer generators, i.e.,
\begin{equation}
\begin{aligned}  \label{stabilizer}
S_{0}=XXX,~~
S_{1}=ZZI,~~
S_{2}=ZIZ,
\end{aligned}
\end{equation}
where
\begin{equation}
\begin{aligned}  \label{Z:X:basis}
Z= \left( \begin{array}{cc} 1 & 0 \\ 0 & -1 \end{array} \right),  ~~
X = \left( \begin{array}{cc} 0 & 1 \\ 1 & 0 \end{array} \right), ~~
I= \left( \begin{array}{cc} 1 & 0 \\ 0 & 1 \end{array} \right),
\end{aligned}
\end{equation}
denote the phase shift, bit flip and no operation acting on the qubit, respectively.
Maneva and Simolin \cite{Maneva:2000:Improved} proposed a multiparty hashing protocol to distill nearly perfect GHZ states by generalizing the quantum XOR operation used in Ref.~\cite{BDSW:1996:Mixed} to the case of multiparty setting. The yield (per input mixed state) in the case of asymptotic data is given by \cite{Maneva:2000:Improved}
\begin{equation}
\begin{aligned}  \label{distillation:yield}
D_{h}=1-\max\{H(e_{b_{1}}),H(e_{b_{2}})\}-H(e_{p}).
\end{aligned}
\end{equation}
Here $H(x)=-x\log _{2}(x)-(1-x)\log _{2}(1-x)$ is the standard binary Shannon entropy function, $e_p$ is the phase shift error rate corresponding to the stabilizer generator $S_{0}$, while $e_{b_{1}}$ and $e_{b_{2}}$ represent the bit flip error rates corresponding to the stabilizer generator $S_{1}$ and $S_{2}$, respectively. One can choose two (classical) random hashing codes, one of which is used to correct bit flip errors and the other one is used to correct phase errors. This can be done by local operation and classical communication with the help of multilateral quantum XOR operations.

Consider the tripartite density matrix $\rho_{ABC}$ which describes the qubit system of Alice, Bob and Charlie \cite{Cirac:1999:se:1,Chen:2007:Multi-pattite}
\begin{equation}
\begin{aligned}  \label{density:matrix:ABC:eq}
\rho_{ABC}=&\lambda_{1}\ket{\Phi_{0}^+}\bra{\Phi_{0}^+}+\lambda_{2}\ket{\Phi_{0}^-}\bra{\Phi_{0}^-}+\lambda_{3}\ket{\Psi_{1}^+}\bra{\Psi_{1}^+}+\lambda_{4}\ket{\Psi_{1}^-}\bra{\Psi_{1}^-}\\
&+\lambda_{5}\ket{\Psi_{2}^+}\bra{\Psi_{2}^+}+\lambda_{6}\ket{\Psi_{2}^-}\bra{\Psi_{2}^-}+\lambda_{7}\ket{\Psi_{3}^+}\bra{\Psi_{3}^+}+\lambda_{8}\ket{\Psi_{3}^-}\bra{\Psi_{3}^-},
\end{aligned}
\end{equation}
where $\sum_{i=1}^8\lambda_{i}=1$. $e_{b_{1}}^Z$ and $e_{b_{2}}^Z$ are defined as the bit flip error rates between Alice and Bob's bits and between Alice and Charlie's bits in $Z$ basis corresponding to the stabilizer generator $S_{1}$ and $S_{2}$, respectively, which can be obtained from Eq.~\eqref{GHZ:States:eq1} and Eq.~\eqref{density:matrix:ABC:eq} as,
\begin{equation}
\begin{aligned}  \label{bit-flip:Z1Z2}
e_{b_{1}}^Z&=\lambda_{3}+\lambda_{4}+\lambda_{5}+\lambda_{6},\\
e_{b_{2}}^Z&=\lambda_{5}+\lambda_{6}+\lambda_{7}+\lambda_{8}.
\end{aligned}
\end{equation}
We employ the bit error rate $e_{b}^{Z}$ to represent the probability that all the bit values of Alice, Bob and Charlie are not the same,
\begin{equation}
\begin{aligned}  \label{bit-flip:Z}
e_{b}^{Z}&=\lambda_{3}+\lambda_{4}+\lambda_{5}+\lambda_{6}+\lambda_{7}+\lambda_{8}.
\end{aligned}
\end{equation}
The phase shift error rate corresponding to the stabilizer generator $S_{0}$ in $Z$ basis can be given by
\begin{equation}
\begin{aligned}  \label{phase-shift:Z}
e_p^Z=\lambda_{2}+\lambda_{4}+\lambda_{6}+\lambda_{8}.
\end{aligned}
\end{equation}
Furthermore, if Alice, Bob and Charlie measure the GHZ state in $X$ basis, the random measurement outcomes will always share a binary correlation $X_{A}=X_{B} \oplus X_{C}$. The bit flip error rate in $X$ basis is the probability of $X_{A}\oplus1 = X_{B}\oplus X_{C}$, while the phase shift error rate in $X$ basis is the probability that the relative phase changes. $X_{A}\in\{0,~1\}$ is the binary data corresponding to the polarization $\{\ket{+},~\ket{-}\}$ of Alice. Therefore, from Eq.~\eqref{GHZ:States:eq1} and Eq.~\eqref{density:matrix:ABC:eq}, the bit flip error rate and phase shift error rate in $X$ basis can be given by
\begin{equation}
\begin{aligned}  \label{bit-flip:X}
e_b^X&=\lambda_{2}+\lambda_{4}+\lambda_{6}+\lambda_{8}=e_p^{Z},\\
e_p^X&=\lambda_{3}+\lambda_{4}+\lambda_{5}+\lambda_{6}+\lambda_{7}+\lambda_{8}=e_{b}^{Z}.
\end{aligned}
\end{equation}

\subsection{B. Post-selected GHZ States}
\label{sec:post-GHZ}
Entanglement purification of GHZ states are closely related to multiparty communication protocols, such as quantum cryptographic conferencing (QCC) \cite{Bose:1998:Multiparticle,Chen:2007:Multi-pattite} and quantum secret sharing (QSS) \cite{Hillery:1999:quantum,Chen:2005:QSS,Gaertner:2007:exp}. The relation between them is that if Alice, Bob and Charlie share almost perfect pure GHZ states, the states will be nearly unentangled with Eve's system, which is the term \emph{monogamy of entanglement} \cite{Terhal:2004}. Therefore, the information leaked to Eve is negligible, and Alice, Bob and Charlie can obtain an information-theoretically secure key by measuring the GHZ states. Thus, the purpose of QCC and QSS protocols can be recognized as a procedure for Alice, Bob and Charlie to share almost perfect GHZ states, which is also the purpose of the entanglement purification protocol. Entanglement purification protocol can be transformed into the quantum error correction protocol \cite{BDSW:1996:Mixed}, while Calderbank-Shor-Steane (CSS) code can be used to prove the security of quantum communication protocols \cite{Shor:2000:Simple,Lo:2001:Sixstate}.
With the important property of CSS code, the error correction procedure for the phase shift error will be decoupled from the error correction procedure for the bit flip error. The quantum error correction can be transformed into classical post-processing, the bit error correction (phase error correction) can be regarded as the classical error correction (privacy amplification).

We use a GHZ-state analyzer \cite{Pan:1999:GHZ} to post-select GHZ states among three legitimate users (Alice, Bob and Charlie). The events can be regarded as the time-reversed GHZ state distribution and measurement. Similar to the security proof of measurement-device-independent (MDI) quantum key distribution \cite{Lo:MIQKD:2012,Lo:1999:unconditional,Shor:2000:Simple}, we suppose that each of Alice, Bob and Charlie has an Einstein-Podolsky-Rosen entangled state which contains one virtual qubit in each of them and the other qubit is sent to the middle node, David. When David performs a successful GHZ-state measurement, the virtual qubit of the legitimate users becomes a GHZ-entangled state, the procedure of which can be then regarded as a multiparty entanglement swapping, as experimentally demonstrated \cite{Lu:2009:GHZ}. Alice, Bob and Charlie can utilize quantum memory to store their virtual qubits. After David announces the events through public channels whether he has obtained a GHZ state and which GHZ state he has received, Alice, Bob and Charlie will measure their virtual qubits.  According to different multiparty quantum communication protocols such as the QCC, QSS and third-man quantum cryptography, the legitimate users perform the corresponding operations to classical post-processing. They can extract secure keys after the processes of basis sift, error correction and privacy amplification, which are all classical procedures. Combined with the decoy sate method \cite{Hwang:2003:Quantum,Lo:2005:Decoy,Wang:2005:Beating}, some practical sources can be used in our schemes for multiparty quantum communication. For instance, weak coherent sources emitted by laser diodes are used in MDI-QCC, weak coherent states with extra classical bit information (the phase post-selection technique) are used in MDI-QSS. Meanwhile, heralded single-photon sources (also called triggered spontaneous parametric down conversion sources) are used in MDI-QSS. Furthermore, we exploit the quantum non-demolition measurement technique \cite{Grangier:QND:1998} to effectively realize a long distribution distance MDI-QSS with weak coherent sources.

\section{II. MDI-quantum cryptographic conferencing} \label{SM:QCC}
When the phases of the weak coherent pulses sent by Alice, Bob and Charlie are fully randomized, the density matrix of the coherent states can be written as
\begin{equation}
\begin{aligned} \rho_{1} = \int_0^{2\pi} \frac{d\theta}{2\pi}
\ket{e^{i\theta}\sqrt{\mu}}\bra{e^{i\theta}\sqrt{\mu}}=e^{-\mu}\sum_{n=0}^\infty \frac{\mu^n}{n!}
\ket{n}\bra{n},\\
\end{aligned}\label{density mateix of coherent state}
\end{equation}%
where $\theta $ and $\mu $ are the phase and intensity of the coherent states, respectively. Then the quantum channel can be considered as a photon number channel \cite{Lo:2005:Decoy}. Note that the multi-photon components are tagged ones whose information will be fully leaked to Eve \cite{GLLP:2004:Security}, the secure key rate of MDI-QCC can be given by
\begin{equation}
\begin{aligned}  \label{MDI-QCC:KeyRate}
R_{QCC}=Q_{v}^{Z}+Q_{111}^{Z}[1-H(e_{111}^{PZ})]- \max \left\{H(E_{\mu\nu\omega}^{ZAB}),H(E_{\mu\nu\omega}^{ZAC})\right\}fQ_{\mu\nu\omega}^{Z},
\end{aligned}
\end{equation}
where $Q_{111}^{Z}=\mu\nu\omega e^{-\mu-\nu-\omega}Y_{111}^{Z}$ is the gain of the single-photon states in $Z$ basis, $Q_{v}^{Z}=e^{-\mu}Q_{0\nu\omega}^{Z}$ is the gain that Alice sends out vacuum state component in $Z$ basis and David obtains a GHZ state measurement result. Here, we assume that Alice's raw key is the reference raw key. For single-photon states, the phase error probability $e_{111}^{PZ}$ in $Z$ basis is equal to the bit error probability $e_{111}^{BX}$ in $X$ basis in the case of asymptotic data according to  Eq.~\eqref{bit-flip:X}, i.e., $e_{111}^{PZ}=e_{111}^{BX}$. $Q_{\mu\nu\omega}^{Z}$ is the overall gain in $Z$ basis and $f$ is the error correction efficiency. $E_{\mu\nu\omega}^{ZAB}$ ($E_{\mu\nu\omega}^{ZAC}$) is the bit flip error rate between Alice's and Bob's (Charlie's) bits in $Z$ basis.

In the following, we will focus on the evolution of the joint quantum states before they enter the detectors. Due to the basis sift in the classical post-processing, we only discuss the case of $ZZZ$ and $XXX$. The joint quantum states of Alice, Bob and Charlie sending out horizontal polarization weak coherent states can be given by
\begin{equation} \label{quantum state in}
\begin{aligned}
&\ket{e^{i\phi_{a}}\sqrt{\mu}}_{H}\ket{e^{i\phi_{b}}\sqrt{\nu}}_{H}\ket{e^{i\phi_{c}}\sqrt{\omega}}_{H},
\end{aligned}
\end{equation}
where $\phi_{a}$, $\phi_{b}$ and $\phi_{c}$ are the overall randomized phases. Then the quantum states arriving at David's GHZ state measurement device (before the quantum states enter the detectors) are given by
\begin{equation} \label{quantum state out}
\begin{aligned}
\ket{e^{i\phi_{b}}\sqrt{\frac{\nu\eta_{b}}{2}}}_{1H}\ket{e^{i\phi_{b}}\sqrt{\frac{\nu\eta_{b}}{2}}}_{1V}\ket{e^{i\phi_{c}}\sqrt{\frac{\omega\eta_{c}}{2}}}_{2H}\ket{e^{i\phi_{c}}\sqrt{\frac{\omega\eta_{c}}{2}}}_{2V}\ket{e^{i\phi_{a}}\sqrt{\frac{\mu\eta_{a}}{2}}}_{3H}\ket{e^{i\phi_{a}}\sqrt{\frac{\mu\eta_{a}}{2}}}_{3V},
\end{aligned}
\end{equation}
where the six detection modes are $1H$, $1V$, $2H$, $2V$, $3H$ and $3V$, respectively. $\eta_{a}$, $\eta_{b}$, $\eta_{c}$ are the overall detection efficiencies of Alice, Bob and Charlie, respectively. Therefore, the detection probabilities for the six threshold single-photon detectors can be written as
\begin{equation} \label{detector probability}
\begin{aligned}
D_{1H}&=D_{1V}=1-(1-p_{d})\exp\left(-\frac{\nu\eta_{b}}{2}\right),~~
D_{2H}=D_{2V}=1-(1-p_{d})\exp\left(-\frac{\omega\eta_{c}}{2}\right),\\
D_{3H}&=D_{3V}=1-(1-p_{d})\exp\left(-\frac{\mu\eta_{a}}{2}\right).
\end{aligned}
\end{equation}

The gain $Q_{HHH}^{\mu\nu\omega\Phi_{0}^+}$ is defined as the probability that Alice, Bob and Charlie send out horizontal polarization weak coherent states with the intensity of $\mu$, $\nu$ and $\omega$, respectively, with David obtaining a successful GHZ state $\ket{\Phi_{0}^{+}}$ measurement event, which is given by
\begin{equation} \label{gain:HHH}
\begin{aligned}
Q_{HHH}^{\mu\nu\omega\Phi_{0}^+}=&\frac{1}{8}\big[D_{1H}D_{2H}D_{3H}(1-D_{1V})(1-D_{2V})(1-D_{3V})+D_{1H}D_{2V}D_{3V}(1-D_{1V})(1-D_{2H})(1-D_{3H})\\
&+D_{1V}D_{2H}D_{3V}(1-D_{1H})(1-D_{2V})(1-D_{3H})+D_{1V}D_{2V}D_{3H}(1-D_{1H})(1-D_{2H})(1-D_{3V})\big],\\
=&\frac{1}{2}(1-p_{d})^3e^{-\frac{x}{2}}\Big[1-(1-p_{d})e^{-{\frac{\mu\eta_{a}}{2}}}\Big]\Big[1-(1-p_{d})e^{-{\frac{\nu\eta_{b}}{2}}}\Big]\Big[1-(1-p_{d})e^{-{\frac{\omega\eta_{c}}{2}}}\Big],\\
\end{aligned}
\end{equation}
where $x=\mu\eta_{a} +\nu\eta_{b}+\omega\eta_{c}$, $1/8$ stands for the probability of a $\ket{HHH}$ polarization when Alice, Bob and Charlie all choose $Z$ basis, $p_{d}$ is the background
count rate.
Due to symmetry, we have
\begin{equation} \label{gain:HHH}
\begin{aligned}
Q_{HHH}^{\mu\nu\omega\Phi_{0}^+}=Q_{HHH}^{\mu\nu\omega\Phi_{0}^-}=Q_{VVV}^{\mu\nu\omega\Phi_{0}^+}=Q_{VVV}^{\mu\nu\omega\Phi_{0}^-}=A.
\end{aligned}
\end{equation}
According to the above procedures, we have
\begin{equation} \label{gain:otherZ}
\begin{aligned}
Q_{HHV}^{\mu\nu\omega\Phi_{0}^+}=Q_{HVV}^{\mu\nu\omega\Phi_{0}^+}=Q_{HHV}^{\mu\nu\omega\Phi_{0}^-}=Q_{HVV}^{\mu\nu\omega\Phi_{0}^-}=&\frac{p_{d}}{2} (1-p_{d})^3  e^{-x}\left(1-p_{d}-e^{\frac{\nu\eta_{b}}{2}}\right) \left(1-p_{d}-e^{\frac{1}{2}(\mu\eta_{a} +\omega\eta_{c})}I_0\left(\sqrt{\mu\eta_{a}\omega\eta_{c}}\right)\right)=B,\\
Q_{VHH}^{\mu\nu\omega\Phi_{0}^+}=Q_{VHV}^{\mu\nu\omega\Phi_{0}^+}=Q_{VHH}^{\mu\nu\omega\Phi_{0}^-}=Q_{VHV}^{\mu\nu\omega\Phi_{0}^-}=&\frac{p_{d}}{2} (1-p_{d})^3  e^{-x}\left(1-p_{d}-e^{\frac{\omega\eta_{c}}{2}}\right) \left(1-p_{d}-e^{\frac{1}{2}(\mu\eta_{a} +\nu\eta_{b})}I_0\left(\sqrt{\mu\eta_{a}\nu\eta_{b}}\right)\right)=C,\\
Q_{HVH}^{\mu\nu\omega\Phi_{0}^+}=Q_{VVH}^{\mu\nu\omega\Phi_{0}^+}=Q_{HVH}^{\mu\nu\omega\Phi_{0}^-}=Q_{VVH}^{\mu\nu\omega\Phi_{0}^-}=&\frac{p_{d}}{2} (1-p_{d})^3  e^{-x}\left(1-p_{d}-e^{\frac{\mu\eta_{a}}{2}}\right) \left(1-p_{d}-e^{\frac{1}{2}(\nu\eta_{b} +\omega\eta_{c})}I_0\left(\sqrt{\nu\eta_{b}\omega\eta_{c}}\right)\right)=D,
\end{aligned}
\end{equation}
where $I_{0}(x)$ is the modified Bessel function of the first kind.

In the same manner, when Alice, Bob and Charlie all choose $X$ basis, we have
\begin{equation} \label{eq9}
\begin{aligned}
Q_{+++}^{\mu\nu\omega\Phi_{0}^+}=Q_{+--}^{\mu\nu\omega\Phi_{0}^+}=Q_{-+-}^{\mu\nu\omega\Phi_{0}^+}=Q_{--+}^{\mu\nu\omega\Phi_{0}^+}
=Q_{++-}^{\mu\nu\omega\Phi_{0}^-}=Q_{+-+}^{\mu\nu\omega\Phi_{0}^-}=Q_{-++}^{\mu\nu\omega\Phi_{0}^-}=Q_{---}^{\mu\nu\omega\Phi_{0}^-}=E,\\
Q_{+++}^{\mu\nu\omega\Phi_{0}^-}=Q_{+--}^{\mu\nu\omega\Phi_{0}^-}=Q_{-+-}^{\mu\nu\omega\Phi_{0}^-}=Q_{--+}^{\mu\nu\omega\Phi_{0}^-}
=Q_{++-}^{\mu\nu\omega\Phi_{0}^+}=Q_{+-+}^{\mu\nu\omega\Phi_{0}^+}=Q_{-++}^{\mu\nu\omega\Phi_{0}^+}=Q_{---}^{\mu\nu\omega\Phi_{0}^+}=F,
\end{aligned}
\end{equation}
and
\begin{equation} \label{gain:X}
\begin{aligned}
Q_{+++}^{\mu\nu\omega\Phi_{0}^+}=&\frac{1}{8}\int_{0}^{2\pi}\int_{0}^{2\pi}\big[F_{1H}F_{2H}F_{3H}(1-F_{1V})(1-F_{2V})(1-F_{3V})+F_{1H}F_{2V}F_{3V}(1-F_{1V})(1-F_{2H})(1-F_{3H})\\
&+F_{1V}F_{2H}F_{3V}(1-F_{1H})(1-F_{2V})(1-F_{3H})+F_{1V}F_{2V}F_{3H}(1-F_{1H})(1-F_{2H})(1-F_{3V})\big]\frac{d\phi}{2\pi} \frac{d\varphi}{2\pi},\\
Q_{+++}^{\mu\nu\omega\Phi_{0}^-}=&\frac{1}{8}\int_{0}^{2\pi}\int_{0}^{2\pi}\big[F_{1H}F_{2H}F_{3V}(1-F_{1V})(1-F_{2V})(1-F_{3H})+F_{1H}F_{2V}F_{3H}(1-F_{1V})(1-F_{2H})(1-F_{3V})\\
&+F_{1V}F_{2H}F_{3H}(1-F_{1H})(1-F_{2V})(1-F_{3V})+F_{1V}F_{2V}F_{3V}(1-F_{1H})(1-F_{2H})(1-F_{3H})\big]\frac{d\phi}{2\pi} \frac{d\varphi}{2\pi},
\end{aligned}
\end{equation}
where $F_{1H}$ is the detection probability of detection mode $1H$, $\phi=\phi_{a}-\phi_{b}$, $\varphi=\phi_{a}-\phi_{c}$, and
\begin{equation} \label{detector probability1}
\begin{aligned}
F_{1H}&=1-(1-p_{d})e^{-{(\frac{\mu\eta_{a}+\nu\eta_{b}}{4}+\frac{\sqrt{\mu\eta_{a}\nu\eta_{b}}}{2}\cos{\phi})}},~~F_{1V}=1-(1-p_{d})e^{-{(\frac{\mu\eta_{a}+\nu\eta_{b}}{4}-\frac{\sqrt{\mu\eta_{a}\nu\eta_{b}}}{2}\cos{\phi})}},\\
F_{2H}&=1-(1-p_{d})e^{-{(\frac{\nu\eta_{b}+\omega\eta_{c}}{4}+\frac{\sqrt{\nu\eta_{b}\omega\eta_{c}}}{2}\cos{(\varphi-\phi)})}},~~F_{2V}=1-(1-p_{d})e^{-{(\frac{\nu\eta_{b}+\omega\eta_{c}}{4}-\frac{\sqrt{\nu\eta_{b}\omega\eta_{c}}}{2}\cos{(\varphi-\phi)})}},\\
F_{3H}&=1-(1-p_{d})e^{-{(\frac{\mu\eta_{a}+\omega\eta_{c}}{4}+\frac{\sqrt{\mu\eta_{a}\omega\eta_{c}}}{2}\cos{\varphi})}}~~F_{3V}=1-(1-p_{d})e^{-{(\frac{\mu\eta_{a}+\omega\eta_{c}}{4}-\frac{\sqrt{\mu\eta_{a}\omega\eta_{c}}}{2}\cos{\varphi})}}.
\end{aligned}
\end{equation}

The overall gain and quantum bit error rates in $Z$ basis can be given by
\begin{equation}
\begin{aligned}  \label{eq1}
Q_{\mu\nu\omega}^{Z} =Q_{\mu\nu\omega}^{CZ}+Q_{\mu\nu\omega}^{EZ}&=Q_{\mu\nu\omega}^{CZAB}+Q_{\mu\nu\omega}^{EZAB}=Q_{\mu\nu\omega}^{CZAC}+Q_{\mu\nu\omega}^{EZAC}=\sum_{n=0}^\infty\sum_{m=0}^\infty\sum_{l=0}^\infty\frac{\mu^n\nu^m\omega^l}{n!m!l!}e^{-\mu-\nu-\omega}Y_{nml}^{Z},\\
E_{\mu\nu\omega}^{Z}Q_{\mu\nu\omega}^{Z}&=e_{d}Q_{\mu\nu\omega}^{CZ}+(1-e_{d})Q_{\mu\nu\omega}^{EZ}=\sum_{n=0}^\infty\sum_{m=0}^\infty\sum_{l=0}^\infty\frac{\mu^n\nu^m\omega^l}{n!m!l!}e^{-\mu-\nu-\omega}e_{nml}^{BZ}Y_{nml}^{Z},\\
E_{\mu\nu\omega}^{ZAB}Q_{\mu\nu\omega}^{Z}&=e_{d}Q_{\mu\nu\omega}^{CZAB}+(1-e_{d})Q_{\mu\nu\omega}^{EZAB},~~~E_{\mu\nu\omega}^{ZAC}Q_{\mu\nu\omega}^{Z}=e_{d}Q_{\mu\nu\omega}^{CZAC}+(1-e_{d})Q_{\mu\nu\omega}^{EZAC},
\end{aligned}
\end{equation}
where $E_{\mu\nu\omega}^{Z}$ is defined as the probability that all the bit values of Alice, Bob and Charlie are not the same in $Z$ basis. $Y_{nml}^{Z}$ ($e_{nml}^{BZ}$) is the yield (bit error rate) in $Z$ basis, given that Alice, Bob and Charlie send out $n$-photon, $m$-photon and $l$-photon pulses, respectively. $Q_{\mu\nu\omega}^{CZ}$ ($Q_{\mu\nu\omega}^{EZ}$) is the total gain of a successful GHZ state measurement when the polarization of the pulses sent by Alice, Bob and Charlie are the same (different) in $Z$ basis, which represents a correct (false) measurement result. $Q_{\mu\nu\omega}^{CZAB}$ ($Q_{\mu\nu\omega}^{EZAB}$) is the total gain of a successful GHZ state measurement when the polarization of the pulses sent by Alice and Bob are the same (different) in $Z$ basis, which represents a correct (false) measurement result. $Q_{\mu\nu\omega}^{CZAC}$ ($Q_{\mu\nu\omega}^{EZAC}$) is the total gain of a successful GHZ state measurement when the polarization of the pulses sent by Alice and Charlie are the same (different) in $Z$ basis, which represents a correct (false) measurement result. $e_{d}$ represents the overall misalignment-error probability of the system. Therefore, we have
\begin{equation}
\begin{aligned}  \label{eq2}
Q_{\mu\nu\omega}^{CZ}=&4A,~~Q_{\mu\nu\omega}^{EZ}=4(B+C+D),~~Q_{\mu\nu\omega}^{CZAB}=4A+2B+2D,\\
Q_{\mu\nu\omega}^{EZAB}=2B+4C+2D&,~~Q_{\mu\nu\omega}^{CZAC}=4A+2C+2D,~~Q_{\mu\nu\omega}^{EZAC}=4B+2C+2D.
\end{aligned}
\end{equation}

The overall gain $Q_{\mu\nu\omega}^{X}$ and quantum bit error rate $E_{\mu\nu\omega}^{X}$ in $X$ basis can be given by
\begin{equation}
\begin{aligned}  \label{eq3}
Q_{\mu\nu\omega}^{X} &=Q_{\mu\nu\omega}^{CX}+Q_{\mu\nu\omega}^{EX}=\sum_{n=0}^\infty\sum_{m=0}^\infty\sum_{l=0}^\infty\frac{\mu^n\nu^m\omega^l}{n!m!l!}e^{-\mu-\nu-\omega}Y_{nml}^{X},\\
E_{\mu\nu\omega}^{X}Q_{\mu\nu\omega}^{X}&=e_{d}Q_{\mu\nu\omega}^{CX}+(1-e_{d})Q_{\mu\nu\omega}^{EX}=\sum_{n=0}^\infty\sum_{m=0}^\infty\sum_{l=0}^\infty\frac{\mu^n\nu^m\omega^l}{n!m!l!}e^{-\mu-\nu-\omega}e_{nml}^{BX}Y_{nml}^{X},\\
\end{aligned}
\end{equation}
where $Y_{nml}^{X}$ ($e_{nml}^{BX}$) is the yield (bit error rate) in $X$ basis, given that Alice, Bob and Charlie send out $n$-photon, $m$-photon and $l$-photon pulses, respectively. $Q_{\mu \nu \omega}^{CX}$ ($Q_{\mu\nu\omega }^{EX}$) is the total gain of a successful GHZ state measurement when the correlation $X_{A}= X_{B}\oplus X_{C}$ ($X_{A}\oplus1= X_{B}\oplus X_{C}$) holds in $X$ basis, which represents a correct (false) measurement result. Thus, we have $Q_{\mu\nu\omega}^{CX}=8E,~Q_{\mu\nu\omega}^{EX}=8F$. Notice that Alice performs a bit flip when Alice, Bob and Charlie all choose
$X$ basis and David obtains the GHZ state $\ket{\Phi_{0}^-}$.

For simplicity, we consider a symmetric scenario that the distances $L$ from Alice, Bob and Charlie to the middle node David are all the same. So $\eta_{a}=\eta_{b}=\eta_{c}=\eta_{d}\times10^{-\beta L/10}$ is the overall efficiency including the channel transmission efficiency $10^{-\beta L/10}$ ($\beta$ is the intrinsic loss coefficient of the standard telecom fiber channel and $L$ is the distance between the legitimate users and David) and the efficiency of the detectors $\eta_{d}$.  We present an analytical estimation method with two decoy states (vacuum+decoy state), here $\mu_{2}=\nu_{2}=\omega_{2}>\mu_{1}=\nu_{1}=\omega_{1}>0$. With the derivation method mentioned in \cite{Xu:2013:practical}, we can calculate the lower bound of $Y_{111}^{ZL}$, $Y_{111}^{XL}$ and the upper bound of $e_{111}^{BXU}$, $e_{111}^{BZU}$, which are given by
\begin{equation}
\begin{aligned}  \label{eq5}
Y_{111}^{ZL}\geq&\frac{1}{\mu_{2}^3\mu_{1}^3(\mu_{2}-\mu_{1})}\Big[\mu_{2}^4\Big(e^{3\mu_{1}}Q_{\mu_{1}\mu_{1}\mu_{1}}^{Z}-e^{2\mu_{1}}Q_{\mu_{1}\mu_{1}0}^{Z}-e^{2\mu_{1}}Q_{\mu_{1}0\mu_{1}}^{Z}-e^{2\mu_{1}}Q_{0\mu_{1}\mu_{1}}^{Z}+e^{\mu_{1}}Q_{\mu_{1}00}^{Z}\\
&+e^{\mu_{1}}Q_{0\mu_{1}0}^{Z}+e^{\mu_{1}}Q_{00\mu_{1}}^{Z}-Q_{000}^{Z}\Big)-\mu_{1}^4\Big(e^{3\mu_{2}}Q_{\mu_{2}\mu_{2}\mu_{2}}^{Z}-e^{2\mu_{2}}Q_{\mu_{2}\mu_{2}0}^{Z}\\
&-e^{2\mu_{2}}Q_{\mu_{2}0\mu_{2}}^{Z}-e^{2\mu_{2}}Q_{0\mu_{2}\mu_{2}}^{Z}+e^{\mu_{2}}Q_{\mu_{2}00}^{Z}+e^{\mu_{2}}Q_{0\mu_{2}0}^{Z}+e^{\mu_{2}}Q_{00\mu_{2}}^{Z}-Q_{000}^{Z}\Big)\Big],
\end{aligned}
\end{equation}
\begin{equation}
\begin{aligned}  \label{eq6}
Y_{111}^{XL}\geq&\frac{1}{\mu_{2}^3\mu_{1}^3(\mu_{2}-\mu_{1})}\Big[\mu_{2}^4\Big(e^{3\mu_{1}}Q_{\mu_{1}\mu_{1}\mu_{1}}^{X}-e^{2\mu_{1}}Q_{\mu_{1}\mu_{1}0}^{X}-e^{2\mu_{1}}Q_{\mu_{1}0\mu_{1}}^{X}-e^{2\mu_{1}}Q_{0\mu_{1}\mu_{1}}^{X}+e^{\mu_{1}}Q_{\mu_{1}00}^{X}\\
&+e^{\mu_{1}}Q_{0\mu_{1}0}^{X}+e^{\mu_{1}}Q_{00\mu_{1}}^{X}-Q_{000}^{X}\Big)-\mu_{1}^4\Big(e^{3\mu_{2}}Q_{\mu_{2}\mu_{2}\mu_{2}}^{X}-e^{2\mu_{2}}Q_{\mu_{2}\mu_{2}0}^{X}\\
&-e^{2\mu_{2}}Q_{\mu_{2}0\mu_{2}}^{X}-e^{2\mu_{2}}Q_{0\mu_{2}\mu_{2}}^{X}+e^{\mu_{2}}Q_{\mu_{2}00}^{X}+e^{\mu_{2}}Q_{0\mu_{2}0}^{X}+e^{\mu_{2}}Q_{00\mu_{2}}^{X}-Q_{000}^{X}\Big)\Big],
\end{aligned}
\end{equation}
\begin{equation}
\begin{aligned}  \label{eq7}
e_{111}^{BXU}\leq&\frac{1}{\mu_{1}^3Y_{111}^{XL}}\Big(e^{3\mu_{1}}E_{\mu_{1}\mu_{1}\mu_{1}}^{X}Q_{\mu_{1}\mu_{1}\mu_{1}}^{X}-e^{2\mu_{1}}E_{\mu_{1}\mu_{1}0}^{X}Q_{\mu_{1}\mu_{1}0}^{X}-e^{2\mu_{1}}E_{\mu_{1}0\mu_{1}}^{X}Q_{\mu_{1}0\mu_{1}}^{X}-e^{2\mu_{1}}E_{0\mu_{1}\mu_{1}}^{X}Q_{0\mu_{1}\mu_{1}}^{X}\\
&+e^{\mu_{1}}E_{\mu_{1}00}^{X}Q_{\mu_{1}00}^{X}+e^{\mu_{1}}E_{0\mu_{1}0}^{X}Q_{0\mu_{1}0}^{X}+e^{\mu_{1}}E_{00\mu_{1}}^{X}Q_{00\mu_{1}}^{X}-E_{000}^{X}Q_{000}^{X}\Big).
\end{aligned}
\end{equation}

\begin{equation}
\begin{aligned}  \label{eq8}
e_{111}^{BZU}\leq&\frac{1}{\mu_{1}^3Y_{111}^{ZL}}\Big(e^{3\mu_{1}}E_{\mu_{1}\mu_{1}\mu_{1}}^{Z}Q_{\mu_{1}\mu_{1}\mu_{1}}^{Z}-e^{2\mu_{1}}E_{\mu_{1}\mu_{1}0}^{Z}Q_{\mu_{1}\mu_{1}0}^{Z}-e^{2\mu_{1}}E_{\mu_{1}0\mu_{1}}^{Z}Q_{\mu_{1}0\mu_{1}}^{Z}-e^{2\mu_{1}}E_{0\mu_{1}\mu_{1}}^{Z}Q_{0\mu_{1}\mu_{1}}^{Z}\\
&+e^{\mu_{1}}E_{\mu_{1}00}^{Z}Q_{\mu_{1}00}^{Z}+e^{\mu_{1}}E_{0\mu_{1}0}^{Z}Q_{0\mu_{1}0}^{Z}+e^{\mu_{1}}E_{00\mu_{1}}^{Z}Q_{00\mu_{1}}^{Z}-E_{000}^{Z}Q_{000}^{Z}\Big).
\end{aligned}
\end{equation}

\section{III. MDI-quantum secret sharing} \label{SM:QCC}
\subsection{A. MDI-QSS with Phase Post-selection Technique}
The MDI-QCC (MDI-QSS) protocol uses the data in $Z$ ($X$) basis to extract secure key. Thus, the secure key rate of MDI-QSS can be given by
\begin{equation}
\begin{aligned}  \label{MDI-QSS:KeyRate}
R_{QSS}=Q_{v}^{X}+Q_{111}^{X}[1-H(e_{111}^{PX})]-Q_{\mu\nu\omega}^{X}fH(E_{\mu\nu\omega}^{X}).
\end{aligned}
\end{equation}
where $Q_{111}^{X}=\mu\nu\omega e^{-\mu-\nu-\omega}Y_{111}^{X}$. In the case of asymptotic data, for single-photon states, the phase error probability in $X$ basis is equal to the bit error probability in $Z$ basis according to Eq.~\eqref{bit-flip:X}, i.e., $e_{111}^{PX}=e_{111}^{BZ}$. $Q_{v}^{X}=e^{-\mu}Q_{0\nu\omega}^{X}$ is the gain that Alice sends out vacuum state component in $X$ basis and David obtains a GHZ state measurement result. $Q_{\mu\nu\omega}^{X}$ ($E_{\mu\nu\omega}^{X}$) is the overall gain (bit error rate) in $X$ basis, which can be directly obtained from the experimental results. Due to that the three parties send out vacuum state, single-photon state and two-photon state in $X$ basis, respectively, David also obtains a GHZ state measurement result and the probability is the same order with that all the three parties send out single-photon states, i.e., $Q_{111}^{X}/2\sim Q_{012}^{X}\sim Q_{021}^{X}\sim Q_{102}^{X}\sim Q_{120}^{X}\sim Q_{201}^{X}\sim Q_{210}^{X} \gg Q_{ijk}^{X}$ for $i+j+k>3$. Therefore, the overall bit error rate in $X$ basis can be written as
\begin{equation}
\begin{aligned}  \label{MDI-QSS:error rate}
E_{\mu\nu\omega}^{X}\sim \frac{6e_{012}Q_{012}^{X}}{Q_{111}^{X}+6Q_{012}^{X}}=37.5\%,
\end{aligned}
\end{equation}
where $e_{012}=50\%$ since the vacuum state carries no bit information. However, the overall quantum bit error rate in $X$ basis is so high that it is virtually impossible to use weak coherent sources to perform MDI-QSS with Eq. \eqref{MDI-QSS:KeyRate}. Fortunately, we can
exploit the extra classical bit information \cite{Ma:Alternative:2012} to extract the raw key with little bit error rate (almost zero) so that we can implement the MDI-QSS with weak coherent sources. With the decoy state method \cite{Hwang:2003:Quantum,Lo:2005:Decoy,Wang:2005:Beating}, the overall phase are randomized over $[0,2\pi)$, which can be divided into $K$ parts in the following form
\begin{equation}
\begin{aligned}  \label{phase randomization}
[0,2\pi)=\bigcup_{k=0}^{K-1}\{[\frac{k\pi}{K},\frac{(k+1)\pi}{K})\cup[\pi+\frac{k\pi}{K},\pi+\frac{(k+1)\pi}{K})\}.
\end{aligned}
\end{equation}
Different regions can be denoted by classical bit information, for example, 3-bit classical information represents $K=8$ phase regions. At the same time that Alice, Bob and Charlie announce their basis, they also announce their overall phase regions. Note that different overall phase regions correspond to different bit error rates, they can extract the raw key with little bit error rate according to phase bit information. Only when their phase regions are chosen the same, the bit error rates will reach the minimum value. Alice, Bob and Charlie only choose the data in the phase region $[0,\frac{\pi}{K})\cup [\pi, \pi+\frac{\pi}{K})$ as the effective raw key. Thus, the gain and bit error rate of post-selection raw key can be written as
\begin{equation}
\begin{aligned}  \label{gain and error}
\widetilde{Q}_{\mu\nu\omega}^{X}=\widetilde{Q}_{\mu\nu\omega}^{CX}+\widetilde{Q}_{\mu\nu\omega}^{EX},~~~~~
\widetilde{E}_{\mu\nu\omega}^{X}\widetilde{Q}_{\mu\nu\omega}^{X}=(1-e_{d})\widetilde{Q}_{\mu\nu\omega}^{EX}+e_{d}\widetilde{Q}_{\mu\nu\omega}^{CX},
\end{aligned}
\end{equation}
where
\begin{equation}
\begin{aligned}  \label{gain phase regions}
\widetilde{Q}_{\mu\nu\omega}^{CX}=&\frac{K}{\pi^3}\int_{0}^{\frac{\pi}{K}}\int_{0}^{\frac{\pi}{K}}\int_{0}^{\frac{\pi}{K}}\big[F_{1H}F_{2H}F_{3H}(1-F_{1V})(1-F_{2V})(1-F_{3V})+F_{1H}F_{2V}F_{3V}(1-F_{1V})(1-F_{2H})(1-F_{3H})\\
&+F_{1V}F_{2H}F_{3V}(1-F_{1H})(1-F_{2V})(1-F_{3H})+F_{1V}F_{2V}F_{3H}(1-F_{1H})(1-F_{2H})(1-F_{3V})\big]d\phi_{a}d\phi_{b}d\phi_{c},\\
\widetilde{Q}_{\mu\nu\omega}^{EX}=&\frac{K}{\pi^3}\int_{0}^{\frac{\pi}{K}}\int_{0}^{\frac{\pi}{K}}\int_{0}^{\frac{\pi}{K}}\big[F_{1H}F_{2H}F_{3V}(1-F_{1V})(1-F_{2V})(1-F_{3H})+F_{1H}F_{2V}F_{3H}(1-F_{1V})(1-F_{2H})(1-F_{3V})\\
&+F_{1V}F_{2H}F_{3H}(1-F_{1H})(1-F_{2V})(1-F_{3V})+F_{1V}F_{2V}F_{3V}(1-F_{1H})(1-F_{2H})(1-F_{3H})\big]d\phi_{a}d\phi_{b}d\phi_{c}.
\end{aligned}
\end{equation}
We assume the gain and bit error rate of single-photon states to be in a uniform distribution over $[0,2\pi)$ \cite{Ma:Alternative:2012}. Therefore, the secure key rate of MDI-QSS with phase post-selection can be given by
\begin{equation}
\begin{aligned} \label{QSS:key:rate:2}
\widetilde{R}_{QSS}\geq &\frac{1}{K^2}Q_{111}^{X}[1-H(e_{111}^{BZ})]-H(\widetilde{E}_{\mu\nu\omega}^{X})f\widetilde{Q}_{\mu\nu\omega}^{X},
\end{aligned}
\end{equation}
where $1/K^2$ represents the probability that all users select the same phase region and we neglect the contribution of vacuum state component that Alice sends out.

In practical experiments, the phase of the transferred signal will drift due to, e.g., temperature and mechanical stress variations on the optical fiber or air disturbance of the free-space channel. Fortunately, the drift of phase will not influence the results of our work, except for MDI-QSS with phase post-selection technique.
The common phase reference is thus required to be shared among all users so that the users can tell which phase region they are. We remark that solving the problem of sharing common phase reference is to tackle long distance phase-stabilization, which is usually difficult and required also in quantum fingerprinting \cite{Arrazola:2014:fingerprinting} and quantum digital signatures \cite{Arrazola:2014:QDS,Dunjko:2014:QDS}.

Here, we suggest a possible way to implement phase compensation over a distance to enable the distribution of the common phase reference among all users. Alice, Bob and Charlie exploit continuous-wave laser sources with the same central wavelength and narrow line-width to generate continuous-wave laser with almost stabilized phases. The amplitude modulator generates reference light and signal light. The reference light is used for phase compensation, while the signal light is used for encoding qubits. When three reference lights with positive $45^{\circ }$ polarization and same intensity enter the GHZ-analyzer, the GHZ-analyzer will unambiguously reveal whether the phases among them are the same or not \cite{andersson:2006:Exp}, i.e., detector D1H and D1V compare the phases between Alice and Bob, detector D2H and D2V compare the phases between Bob and Charlie, detector D3H and D3V compare the phases between Alice and Charlie, respectively. Thus, with the detection results corresponding to reference light, one can realize the phase compensation, resulting in a common phase reference among all users. Considering the scattering effects in fiber, the reference light should not be too strong, so as to reduce the detrimental scattering effects. Another approach could be to use wavelength division multiplexing with the frequency of reference light less than that of signal light so that the detrimental scattered photons can be filtered out. As seen from Fig.~1 in the Supplemental Material of Ref  \cite{Liu:2013:Exp}, practically the phase drift is about $30\pi$ per second for 100 km standard single-mode fiber (SMF-28), so rapid feedback algorithm is necessary for implementing long distance phase compensation. There are some rapid feedback algorithms realizing phase-stabilization for several kilometers \cite{cho:2009:stabilization,Liu:2012:CQC,cuevas:2013:long}. However, successfully accomplishing long distance (100 km) phase-stabilization is still challenging under current technology.

It should be noted that the inclusion of phase post-selection complicates the security analysis, as pointed out in the context of device-independent QKD \cite{ma:2012:improved} or MDI-QKD \cite{Ma:Alternative:2012}. The rigorous security of MDI-QSS with weak coherent states and phase post-selection thus needs more investigations, too.

\subsection{B. MDI-QSS with Heralded Single-photon Sources}
Except for the phase post-selection technique, we propose another two methods to perform MDI-QSS: the triggered spontaneous parametric down conversion sources, or the conventional weak coherent state sources together with the quantum non-demolition measurement technique. Instead of taking advantage of weak coherent states which are divided into two independent states after passing through a beam splitter, we use another universal method to process the joint quantum state evolution, which can also be used for any photon-number distribution (including coherent states) of the sources. That is, we use the heralded single-photon sources (also called triggered spontaneous parametric down-conversion sources) to perform MDI-QSS. Similarly to the above symmetric scenario, $\eta=\eta_{a}=\eta_{b}=\eta_{c}=\eta_{d}\times10^{-\beta L/10}$. The quantum states coming from the heralded single-photon sources can be written as
\begin{equation}
\begin{aligned}  \label{SPDC}
\ket{\Psi}=(\cosh\chi^{-1})\sum_{n=0}^{\infty}(\tanh\chi)^n\ket{n,n}.
\end{aligned}
\end{equation}
We assume that the intensity of the sources is given by $\mu=\sinh^2\chi$ and the heralded single-photon sources always send out photon pairs. Therefore, the photon number of two modes are always the same. The probability to get an $n$-photon pair is
\begin{equation}
\begin{aligned}  \label{distirbution}
P(n)=\frac{\mu^n}{(1+\mu)^{n+1}}.
\end{aligned}
\end{equation}
After triggering out one of the photon pairs, the density matrix of the other mode after phase randomization can then be given by \cite{Lutkenhaus:2000:security}
\begin{equation}
\begin{aligned}  \label{SPDC:densitymatrix}
\rho_{2}=\frac{1}{P_{c}}\sum_{n=0}^{\infty}\frac{\mu^n}{(1+\mu)^{n+1}}[1-(1-p_{d})(1-\eta_{d})^n]\ket{n}\bra{n}=\sum_{n=0}^{\infty}P_{n}(\mu)\ket{n}\bra{n},
\end{aligned}
\end{equation}
where $P_{c}=(\mu\eta_{d}+p_{d})/(1+\mu\eta_{d})$ is the post-selection probability given that one triggered mode leads to the click of the threshold single-photon detector.

We consider the joint quantum states when Alice and Bob send out $i$-photon and $j$-photon states with horizontal polarization, respectively, while Charlie sends out $k$-photon state with vertical polarization. The joint quantum states can be written as
\begin{equation} \label{iniHjHkH}
\begin{aligned}
\ket{\Psi}_{in}^{HHV}=\ket{n}_H\ket{m}_H\ket{l}_V=\frac{(a_{1H}^\dag)^n}{\sqrt{n!}}\frac{(a_{2H}^\dag)^m}{\sqrt{m!}}\frac{(a_{3V}^\dag)^l}{\sqrt{l!}}\ket{0}.
\end{aligned}
\end{equation}
The joint quantum states before entering the detectors can be given by
\begin{equation} \label{outiHjHkH}
\begin{aligned}
\ket{\Psi}_{out}^{HHV}=\sum_{p=0}^{n+l}\sum_{s=0}^{m}\sum_{t=0}^{l}\frac{(-1)^{l-t}C_{n}^{p-t}C_{m}^{s}C_{l}^{t}}{\sqrt{2^{n+m+l}n!m!l!}}\sqrt{p!s!(n+l-p)!(m-s)!(l-t)!}\ket{s}_{1H}\ket{m-s}_{1V}\ket{0}_{2H}\ket{0}_{2V}\ket{p}_{3H}\ket{n+l-p}_{3V},
\end{aligned}
\end{equation}
where $\ket{\Psi}_{out}^{HHV}$ denotes the superpositions of orthogonal states $\ket{s}_{1H}\ket{m-s}_{1V}\ket{0}_{2H}\ket{0}_{2V}\ket{p}_{3H}\ket{n+l-p}_{3V}$.
Therefore, the gain $Q_{HHV}^{\mu\nu\omega\Phi_{0}^+}$ and the yield $Y_{nml}^{HHV\Phi_{0}^+}$ can be written as
\begin{equation} \label{QHH}
\begin{aligned}
Q_{HHV}^{\mu\nu\omega\Phi_{0}^+} =&\frac{1}{8}\sum_{n=0}^{\infty}\sum_{m=0}^{\infty}\sum_{l=0}^{\infty}P_{\mu}(n)P_{\nu}(m)P_{\omega}(l)Y_{nml}^{HHV\Phi_{0}^+},\\
Y_{nml}^{HHV\Phi_{0}^+}=&\sum_{p=0}^{n+l}\sum_{s=0}^{m}\big[G_{1H}G_{2H}G_{3H}(1-G_{1V})(1-G_{2V})(1-G_{3V})+G_{1H}G_{2V}G_{3V}(1-G_{1V})(1-G_{2H})(1-G_{3H})\\
&+G_{1V}G_{2H}G_{3V}(1-G_{1H})(1-G_{2V})(1-G_{3H})+G_{1V}G_{2V}G_{3H}(1-G_{1H})(1-G_{2H})(1-G_{3V})\big]P_{nml}^{HHV},
\end{aligned}
\end{equation}
where $P_{nml}^{HHV}$ is the probability of obtaining the quantum state $\ket{s}_{1H}\ket{m-s}_{1V}\ket{0}_{2H}\ket{0}_{2V}\ket{p}_{3H}\ket{n+l-p}_{3V}$, $G_{1H}$ is the detection probability of detector mode $1H$, and
\begin{equation} \label{PnmlHHH}
\begin{aligned}
P_{nml}^{HHV}&=\left|\sum_{t=0}^l\frac{(-1)^{l-t}C_{n}^{p-t}C_{m}^{s}C_{l}^{t}}{\sqrt{2^{n+m+l}n!m!l!}}\sqrt{p!s!(n+l-p)!(m-s)!}\right|^2,\\
G_{1H}&=1-(1-p_{d})(1-\eta)^s,~G_{1V}=1-(1-p_{d})(1-\eta)^{m-s}, ~G_{2H}=p_{d}, \\
G_{2V}&=p_{d}, ~G_{3H}=1-(1-p_{d})(1-\eta)^{p}, ~G_{3V}=1-(1-p_{d})(1-\eta)^{n+l-p}.
\end{aligned}
\end{equation}
The above methods can also be extended to cases of other polarizations.

Combining Eqs.~\eqref{eq1},~\eqref{eq2},~\eqref{eq3} with Eq.~\eqref{SPDC:densitymatrix}, we will obtain $Q_{\mu\mu\mu}^X$ and  $E_{\mu\mu\mu}^X$ under the heralded single-photon sources. Similar to Eqs.~\eqref{eq5},~\eqref{eq6},~\eqref{eq7}, we can obtain the lower bound of $Y_{111}^{XL}$, $Y_{111}^{ZL}$ and the upper bound of $e_{111}^{BZU}$,
\begin{equation}
\begin{aligned}  \label{YieldX2}
Y_{111}^{XL}\geq&\frac{1}{P_{1}^{2}(\mu_{2})P_{1}^2(\mu_{1})\big[P_{2}(\mu_{2})P_{1}(\mu_{1})-P_{2}(\mu_{1})P_{1}(\mu_{2})\big]}\Big[P_{1}^2(\mu_{2})P_{2}(\mu_{2})\Big(Q_{\mu_{1}\mu_{1}\mu_{1}}^{X}-P_{0}(\mu_{1})Q_{\mu_{1}\mu_{1}0}^{X}-P_{0}(\mu_{1})Q_{\mu_{1}0\mu_{1}}^{X}\\
&-P_{0}(\mu_{1})Q_{0\mu_{1}\mu_{1}}^{X}+P_{0}^2(\mu_{1})Q_{\mu_{1}00}^{X}+P_{0}^2(\mu_{1})Q_{0\mu_{1}0}^{X}+P_{0}^2(\mu_{1})Q_{00\mu_{1}}^{X}-P_{0}^3(\mu_{1})Q_{000}^{X}\Big)-P_{1}^2(\mu_{1})P_{2}(\mu_{1})\Big(Q_{\mu_{2}\mu_{2}\mu_{2}}^{X}\\
&-P_{0}(\mu_{2})Q_{\mu_{2}\mu_{2}0}^{X}-P_{0}(\mu_{2})Q_{\mu_{2}0\mu_{2}}^{X}-P_{0}(\mu_{2})Q_{0\mu_{2}\mu_{2}}^{X}+P_{0}^2(\mu_{2})Q_{\mu_{2}00}^{X}+P_{0}^2(\mu_{2})Q_{0\mu_{2}0}^{X}+P_{0}^2(\mu_{2})Q_{00\mu_{2}}^{X}-P_{0}^3(\mu_{2})Q_{000}^{X}\Big)\Big],\\
\end{aligned}
\end{equation}

\begin{equation}
\begin{aligned}  \label{YieldZ2}
Y_{111}^{ZL}\geq&\frac{1}{P_{1}^{2}(\mu_{2})P_{1}^2(\mu_{1})\big[P_{2}(\mu_{2})P_{1}(\mu_{1})-P_{2}(\mu_{1})P_{1}(\mu_{2})\big]}\Big[P_{1}^2(\mu_{2})P_{2}(\mu_{2})\Big(Q_{\mu_{1}\mu_{1}\mu_{1}}^{Z}-P_{0}(\mu_{1})Q_{\mu_{1}\mu_{1}0}^{Z}-P_{0}(\mu_{1})Q_{\mu_{1}0\mu_{1}}^{Z}\\
&-P_{0}(\mu_{1})Q_{0\mu_{1}\mu_{1}}^{Z}+P_{0}^2(\mu_{1})Q_{\mu_{1}00}^{Z}+P_{0}^2(\mu_{1})Q_{0\mu_{1}0}^{Z}+P_{0}^2(\mu_{1})Q_{00\mu_{1}}^{Z}-P_{0}^3(\mu_{1})Q_{000}^{Z}\Big)-P_{1}^2(\mu_{1})P_{2}(\mu_{1})\Big(Q_{\mu_{2}\mu_{2}\mu_{2}}^{Z}\\
&-P_{0}(\mu_{2})Q_{\mu_{2}\mu_{2}0}^{Z}-P_{0}(\mu_{2})Q_{\mu_{2}0\mu_{2}}^{Z}-P_{0}(\mu_{2})Q_{0\mu_{2}\mu_{2}}^{Z}+P_{0}^2(\mu_{2})Q_{\mu_{2}00}^{Z}+P_{0}^2(\mu_{2})Q_{0\mu_{2}0}^{Z}+P_{0}^2(\mu_{2})Q_{00\mu_{2}}^{Z}-P_{0}^3(\mu_{2})Q_{000}^{Z}\Big)\Big],\\
\end{aligned}
\end{equation}
\begin{equation}
\begin{aligned}  \label{errorXU}
e_{111}^{BZU}\leq&\frac{1}{P_{1}^3(\mu_{1})Y_{111}^{XL}}\Big(E_{\mu_{1}\mu_{1}\mu_{1}}^{Z}Q_{\mu_{1}\mu_{1}\mu_{1}}^{Z}-P_{0}(\mu_{1})E_{\mu_{1}\mu_{1}0}^{Z}Q_{\mu_{1}\mu_{1}0}^{Z}-P_{0}(\mu_{1})E_{\mu_{1}0\mu_{1}}^{Z}Q_{\mu_{1}0\mu_{1}}^{Z}\\
&-P_{0}(\mu_{1})E_{0\mu_{1}\mu_{1}}^{Z}Q_{0\mu_{1}\mu_{1}}^{Z}+P_{0}^2(\mu_{1})E_{\mu_{1}00}^{Z}Q_{\mu_{1}00}^{Z}+P_{0}^2(\mu_{1})Q_{0\mu_{1}0}^{Z}Q_{0\mu_{1}0}^{Z}+P_{0}^2(\mu_{1})E_{00\mu_{1}}^{Z}Q_{00\mu_{1}}^{Z}-P_{0}^{3}(\mu_{1})E_{000}^{Z}Q_{000}^{Z}\Big).
\end{aligned}
\end{equation}

\subsection{C. MDI-QSS with Quantum Non-demolition Measurement Technique}
\begin{figure}[tbh]
\centering
\resizebox{9cm}{!}{\includegraphics{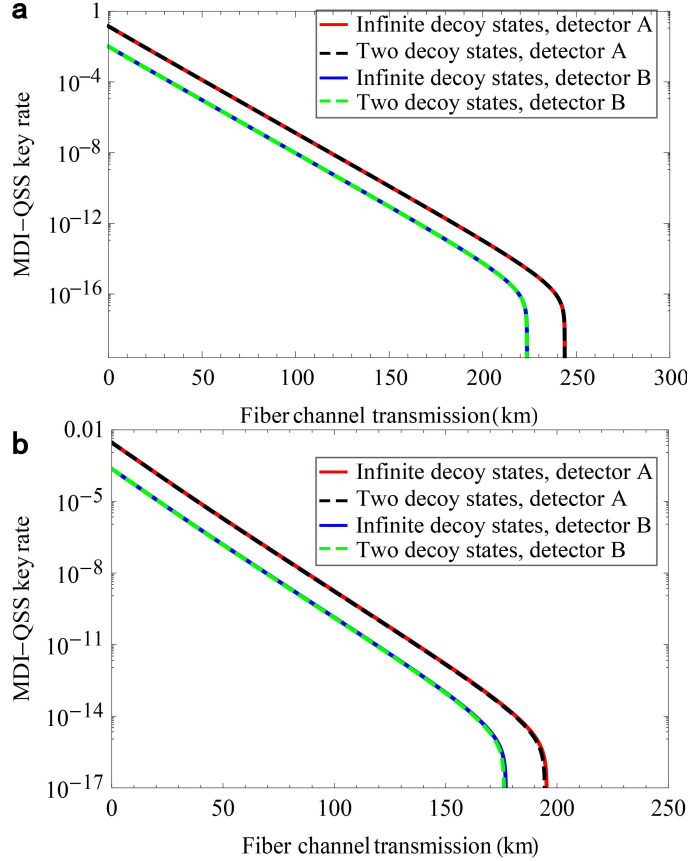}}
\caption{(color online) Lower bound on the secure key rates versus fiber
channel transmission. \textbf{a}, MDI-QSS with heralded single-photon sources.
\textbf{b}, MDI-QSS with weak coherent sources based on quantum non-demolition measurement technique. We show the simulation results of infinite decoy states and two decoy states with detector A (B) of detection efficiency $93\%$ ($40\%$%
), respectively. The overall misalignment-error probability $e_{d}$ of the system is $1.5\%$. The phase-randomized heralded single photon sources are used for MDI-QSS. The intensity of the signal state (one decoy state) is
$5\times10^{-3}$ ($5\times10^{-4}$), while the other decoy state is a vacuum
state.The phase-randomized weak coherent sources are used for MDI-QSS aided by quantum non-demolition
measurement technique. The intensity of the signal state
(one decoy state) is $0.4$ ($0.005$), while the other decoy state
is a vacuum state. }
\label{Fig:4}
\end{figure}

In this subsection, we perform MDI-QSS with weak coherent states by employing quantum non-demolition measurement technique.
The density matrix of phase randomized weak coherent sources after channel transmission can be written as
\begin{equation}
\begin{aligned}
\rho_{3}=e^{-\mu\eta_{t}}\sum_{n=0}^\infty \frac{(\mu\eta_{t})^n}{n!}
\ket{n}\bra{n},\\
\end{aligned}\label{density mateix of coherent state2}
\end{equation}
where the efficiency of channel transmission $\eta_{t}=10^{-\beta L/10}$.
David performs quantum non-demolition measurement on the three incoming pulses from Alice, Bob and Charlie before the pulses enter the GHZ state measurement device. Only when all the photon numbers of the three incoming pulses are no more than one, David will thereafter make a GHZ state measurement.
Therefore, the gain $Q_{HHV}^{\mu\nu\omega\Phi_{0}^+}$ and the yield $Y_{nml}^{HHV\Phi_{0}^+}$ can be written as
\begin{equation} \label{QHH2}
\begin{aligned}
Q_{HHV}^{\mu\nu\omega\Phi_{0}^+} =&\frac{1}{8}\sum_{n=0}^{1}\sum_{m=0}^{1}\sum_{l=0}^{1}e^{-\mu \eta_{t}-\nu\eta_{t}-\omega\eta_{t}}\frac{(\mu\eta_{t})^n}{n!}\frac{(\nu\eta_{t})^m}{m!}\frac{(\omega\eta_{t})^l}{l!}Y_{nml}^{HHV\Phi_{0}^+},\\
Y_{nml}^{HHV\Phi_{0}^+}=&\sum_{p=0}^{n+l}\sum_{s=0}^{m}\big[K_{1H}K_{2H}K_{3H}(1-K_{1V})(1-K_{2V})(1-K_{3V})+K_{1H}K_{2V}K_{3V}(1-K_{1V})(1-K_{2H})(1-K_{3H})\\
&+K_{1V}K_{2H}K_{3V}(1-K_{1H})(1-K_{2V})(1-K_{3H})+K_{1V}K_{2V}K_{3H}(1-K_{1H})(1-K_{2H})(1-K_{3V})\big]P_{nml}^{HHV},
\end{aligned}
\end{equation}
where $K_{1H}$ is the detection probability of detection mode $1H$, and
\begin{equation} \label{PnmlHHH}
\begin{aligned}
P_{nml}^{HHV}&=\left|\sum_{t=0}^l\frac{(-1)^{l-t}C_{n}^{p-t}C_{m}^{s}C_{l}^{t}}{\sqrt{2^{n+m+l}n!m!l!}}\sqrt{p!s!(n+l-p)!(m-s)!}\right|^2,\\
K_{1H}&=1-(1-p_{d})(1-\eta_{d})^s,~K_{1V}=1-(1-p_{d})(1-\eta_{d})^{m-s}, ~K_{2H}=p_{d}, \\
K_{2V}&=p_{d}, ~K_{3H}=1-(1-p_{d})(1-\eta_{d})^{p}, ~K_{3V}=1-(1-p_{d})(1-\eta_{d})^{n+l-p}.
\end{aligned}
\end{equation}
The above methods can also be extended to cases of other polarizations. Similarly to the procedure above, one can calculate the parameters of Eq.~\eqref{MDI-QSS:KeyRate}. With the above two methods, we can obtain the numerical simulation results of the secure key rates of MDI-QSS (see Fig.~\ref{Fig:4}).

\section{IV. Mermin's Inequality} \label{value}

For tripartite systems, each particle is measured by Alice, Bob and Charlie with two bases (settings), local hidden-variable theories must obey Mermin's inequality \cite{Mermin:1990:Inequality}
\begin{equation}
\begin{aligned}  \label{Mermin's:inequality}
M={\langle XXX \rangle}-{\langle XYY \rangle}-{\langle YXY \rangle}-{\langle YYX \rangle}\leq2,
\end{aligned}
\end{equation}
where $M$ is the Mermin value, and
\begin{equation}
\begin{aligned}  \label{X:Y:basis}
X = \left( \begin{array}{cc} 0 & 1 \\ 1 & 0 \end{array} \right), ~~
Y= \left( \begin{array}{cc} 0 & -i \\ i & 0 \end{array} \right).
\end{aligned}
\end{equation}
The Mermin value can reach the maximal value of $4$ given that the tripartite GHZ states are measured under the ideal circumstance, e.g., for
\begin{equation}
\begin{aligned}  \label{GHZ}
\ket{\Phi_{0}^{+}}=\frac{1}{\sqrt{2}}\left(\ket{HHH}+\ket{VVV}\right).
\end{aligned}
\end{equation}
Here, we combine the decoy-state method with weak coherent state sources to estimate the Mermin value of our post-selected GHZ states,
\begin{equation}
\begin{aligned}  \label{GHZ}
M_{111}^{\Phi_{0}^+}&={\langle XXX \rangle}_{111}^{\Phi_{0}^+}-{\langle XYY \rangle}_{111}^{\Phi_{0}^+}-{\langle YXY \rangle}_{111}^{\Phi_{0}^+}-{\langle YYX \rangle}_{111}^{\Phi_{0}^+},
\end{aligned}
\end{equation}
where ${\langle XXX\rangle}_{111}^{\Phi_{0}^+}$ is the expectation value of the GHZ state solely contributed by the single-photon state components, which results from the successful projection into the GHZ state $\ket{\Phi_{0}^{+}}$, given that Alice, Bob and Charlie send out the quantum states of $X$ basis. The expectation value of ${\langle XXX\rangle}_{111}^{\Phi_{0}^+}$ is given by
\begin{equation}
\begin{aligned}
{\langle XXX\rangle}_{111}^{\Phi_{0}^+}=(1-2e_{d})\frac{Y_{+++}^{111\Phi_{0}^{+}}+Y_{+--}^{111\Phi_{0}^{+}}+Y_{-+-}^{111\Phi_{0}^{+}}+Y_{--+}^{111\Phi_{0}^{+}}-Y_{++-}^{111\Phi_{0}^{+}}-Y_{+-+}^{111\Phi_{0}^{+}}-Y_{-++}^{111\Phi_{0}^{+}}-Y_{---}^{111\Phi_{0}^{+}}}
{Y_{+++}^{111\Phi_{0}^{+}}+Y_{+--}^{111\Phi_{0}^{+}}+Y_{-+-}^{111\Phi_{0}^{+}}+Y_{--+}^{111\Phi_{0}^{+}}+Y_{++-}^{111\Phi_{0}^{+}}+Y_{+-+}^{111\Phi_{0}^{+}}+Y_{-++}^{111\Phi_{0}^{+}}+Y_{---}^{111\Phi_{0}^{+}}}.
\end{aligned}  \label{XXX}
\end{equation}
With weak coherent state sources, the gain $Q_{+++}^{\mu\nu\omega\Phi_{0}^+}$ and $Q_{---}^{\mu\nu\omega\Phi_{0}^+}$ can be written as
\begin{equation}
\begin{aligned}  \label{eq8}
Q_{+++}^{\mu\nu\omega\Phi_{0}^+}=\sum_{n=0}^\infty\sum_{m=0}^\infty\sum_{l=0}^\infty\frac{\mu^n\nu^m\omega^l}{n!m!l!}e^{-\mu-\nu-\omega}Y_{+++}^{nml\Phi_{0}^+},~Q_{---}^{\mu\nu\omega\Phi_{0}^+}=\sum_{n=0}^\infty\sum_{m=0}^\infty\sum_{l=0}^\infty\frac{\mu^n\nu^m\omega^l}{n!m!l!}e^{-\mu-\nu-\omega}Y_{---}^{nml\Phi_{0}^+},
\end{aligned}
\end{equation}
where $Y_{+++}^{nml\Phi_{0}^+}$ ($Y_{---}^{nml\Phi_{0}^+}$) is the yield given that Alice, Bob and Charlie send out $n$-photon state, $m$-photon state and $l$-photon state with $\ket{+}$ ($\ket{-}$) polarization, respectively. Thus we can obtain the lower (upper) bound of $Y_{+++}^{111\Phi_{0}^+L}$ ($Y_{+++}^{111\Phi_{0}^+U}$ and $Y_{---}^{111\Phi_{0}^+U}$) in the following,
\begin{equation}
\begin{aligned}  \label{yield1}
Y_{+++}^{111\Phi_{0}^+L}\geq&\frac{1}{\mu_{2}^3\mu_{1}^3(\mu_{2}-\mu_{1})}\Big[\mu_{2}^4\Big(e^{3\mu_{1}}Q_{+++}^{\mu_{1}\mu_{1}\mu_{1}\Phi_{0}^+}-e^{2\mu_{1}}Q_{+++}^{\mu_{1}\mu_{1}0\Phi_{0}^+}-e^{2\mu_{1}}Q_{+++}^{\mu_{1}0\mu_{1}\Phi_{0}^+}-e^{2\mu_{1}}Q_{+++}^{0\mu_{1}\mu_{1}\Phi_{0}^+}+e^{\mu_{1}}Q_{+++}^{\mu_{1}00\Phi_{0}^+}\\
&+e^{\mu_{1}}Q_{+++}^{0\mu_{1}0\Phi_{0}^+}+e^{\mu_{1}}Q_{+++}^{00\mu_{1}\Phi_{0}^+}-Q_{+++}^{000\Phi_{0}^+}\Big)-\mu_{1}^4\Big(e^{3\mu_{2}}Q_{+++}^{\mu_{2}\mu_{2}\mu_{2}\Phi_{0}^+}-e^{2\mu_{2}}Q_{+++}^{\mu_{2}\mu_{2}0\Phi_{0}^+}\\
&-e^{2\mu_{2}}Q_{+++}^{\mu_{2}0\mu_{2}\Phi_{0}^+}-e^{2\mu_{2}}Q_{+++}^{0\mu_{2}\mu_{2}\Phi_{0}^+}+e^{\mu_{2}}Q_{+++}^{\mu_{2}00\Phi_{0}^+}+e^{\mu_{2}}Q_{+++}^{0\mu_{2}0\Phi_{0}^+}+e^{\mu_{2}}Q_{+++}^{00\mu_{2}\Phi_{0}^+}-Q_{+++}^{000\Phi_{0}^+}\Big)\Big],\\
\end{aligned}
\end{equation}
\begin{equation}
\begin{aligned}  \label{yield2}
Y_{+++}^{111\Phi_{0}^+U}\leq&\frac{1}{\mu_{1}^3}\Big(e^{3\mu_{1}}Q_{+++}^{\mu_{1}\mu_{1}\mu_{1}\Phi_{0}^+}-e^{2\mu_{1}}Q_{+++}^{\mu_{1}\mu_{1}0\Phi_{0}^+}-e^{2\mu_{1}}Q_{+++}^{\mu_{1}0\mu_{1}\Phi_{0}^+}-e^{2\mu_{1}}Q_{+++}^{0\mu_{1}\mu_{1}\Phi_{0}^+}+e^{\mu_{1}}Q_{+++}^{\mu_{1}00\Phi_{0}^+}\\
&+e^{\mu_{1}}Q_{+++}^{0\mu_{1}0\Phi_{0}^+}+e^{\mu_{1}}Q_{+++}^{00\mu_{1}\Phi_{0}^+}-Q_{+++}^{000\Phi_{0}^+}\Big).
\end{aligned}
\end{equation}
\begin{equation}
\begin{aligned}  \label{yield3}
Y_{---}^{111\Phi_{0}^+U}\leq&\frac{1}{\mu_{1}^3}\Big(e^{3\mu_{1}}Q_{---}^{\mu_{1}\mu_{1}\mu_{1}\Phi_{0}^+}-e^{2\mu_{1}}Q_{---}^{\mu_{1}\mu_{1}0\Phi_{0}^+}-e^{2\mu_{1}}Q_{---}^{\mu_{1}0\mu_{1}\Phi_{0}^+}-e^{2\mu_{1}}Q_{---}^{0\mu_{1}\mu_{1}\Phi_{0}^+}+e^{\mu_{1}}Q_{---}^{\mu_{1}00\Phi_{0}^+}\\
&+e^{\mu_{1}}Q_{---}^{0\mu_{1}0\Phi_{0}^+}+e^{\mu_{1}}Q_{---}^{00\mu_{1}\Phi_{0}^+}-Q_{---}^{000\Phi_{0}^+}\Big).
\end{aligned}
\end{equation}
From Eq.~\eqref{eq9} and Eq.~\eqref{eq8}, we have
\begin{equation}
\begin{aligned}  \label{yield3}
Y_{+++}^{111\Phi_{0}^+}=Y_{+--}^{111\Phi_{0}^{+}}=Y_{-+-}^{111\Phi_{0}^{+}}=Y_{--+}^{111\Phi_{0}^{+}},\\
Y_{---}^{111\Phi_{0}^+}=Y_{+-+}^{111\Phi_{0}^{+}}=Y_{-++}^{111\Phi_{0}^{+}}=Y_{++-}^{111\Phi_{0}^{+}}.
\end{aligned}
\end{equation}
The lower bound of ${\langle XXX\rangle}_{111}^{\Phi_{0}^+}$ can be given by
\begin{equation}
\begin{aligned}
{\langle XXX\rangle}_{111}^{\Phi_{0}^+L}=(1-2e_{d})\frac{Y_{+++}^{111\Phi_{0}^{+}L}-Y_{---}^{111\Phi_{0}^{+}U}}
{Y_{+++}^{111\Phi_{0}^{+}U}+Y_{---}^{111\Phi_{0}^{+}U}}.
\end{aligned}  \label{XXX}
\end{equation}

Similar to the above methods, we have the expectation values of ${\langle XYY \rangle}_{111}^{\Phi_{0}^+}$, ${\langle YXY \rangle}_{111}^{\Phi_{0}^+}$ and ${\langle YYX \rangle}_{111}^{\Phi_{0}^+}$ as follows,
\begin{equation}
\begin{aligned}  \label{GHZ}
{\langle XXX \rangle}_{111}^{\Phi_{0}^+}=-{\langle XYY \rangle}_{111}^{\Phi_{0}^+}=-{\langle YXY \rangle}_{111}^{\Phi_{0}^+}=-{\langle YYX \rangle}_{111}^{\Phi_{0}^+}.
\end{aligned}
\end{equation}
Therefore, the lower bound of the Mermin value can be given by
\begin{equation}
\begin{aligned}  \label{Lvalue}
M_{111}^{\Phi_{0}^+L}=4{\langle XXX\rangle}_{111}^{\Phi_{0}^+L}=4(1-2e_{d})\frac{Y_{+++}^{111\Phi_{0}^{+}L}-Y_{---}^{111\Phi_{0}^{+}U}}
{Y_{+++}^{111\Phi_{0}^{+}U}+Y_{---}^{111\Phi_{0}^{+}U}}.
\end{aligned}
\end{equation}

\section{V. Mermin's Three-particle Version of the Kochen-Specker Theorem} \label{value}

The usual GHZ experiment goes by creating a (post-selected) GHZ entangled state and then sending each particle in the GHZ entanglement over a distance to Alice, Bob and Charlie, each of whom measures the received particle along a randomly chosen basis (either $X$ basis or $Y$ basis). Each of measured values for each observer should have a predetermined value and as such, Mermin's inequality like Eq. (4) in the main text necessarily follows, as required by local realism, which can be ruled out by performing the actual GHZ experiment.

However, the protocol for demonstrating the violation of Mermin's inequality is in some sense the time-reversed GHZ experiment, where the state preparations replace the state measurements in the usual GHZ test and the GHZ-entangled state is measured at the end of each run of the experiment, rather than prepared at the beginning of each run. The interpretation of such a time-reversed GHZ experiment and, in particular, its relevance to the test of (local) realism have never been considered in the literature to the best of our knowledge and are thus interesting in its own right.

While it is beyond the scope of the main text of the present paper to clarify the point, here we would like to argue that the proposed  time-reversed GHZ experiment enables the test of a particular form of the Kochen-Specker theorem \cite{kochen:1967:problem} as proposed by Mermin \cite{Mermin:1990:Simple}. The usual Bell theorem (Bell's inequalities and the GHZ theorem) has three independent assumptions  \cite{scheidl:2010:violation}: locality, realism and freedom of choices (namely, the experimental setting choices are truly random and free). However, in the proposed time-reversed GHZ experiment, we can suppose that Alice, Bob and Charlie prepare their own single-photon states randomly either in the $X$ basis or in the $Y$ basis; as a proof-of-principle argument, we do not use the weak coherent light sources to avoid the experimental complication caused by the non-ideal light sources. The three single photons are then subject to the GHZ measurement at David's station. The measurements and the preparations of these single photons cannot be spacelike-separated. Then we immediately see that the proposed time-reversed GHZ experiment does not test local realism. Instead, we argue that what it actually tests is the Kochen-Specker theorem as proposed by Mermin for the case of eight-dimensional space of three spins/qubits \cite{Mermin:1990:Simple}.

The Kochen-Specker theorem states that quantum mechanical predictions for any systems of dimensions 3 or higher cannot be reproduced by noncontextual hidden-variable theories that assume the measurement results to be predetermined and independent of other compatible measurements.
In Mermin's argument of the Kochen-Specker theorem, one makes use of a set of the operator identities:
\begin{equation}
\begin{aligned}  \label{op-identity}
X_{1}X_{2}X_{3}\cdot X_{1}\cdot X_{2} \cdot X_{3}&=1,\\
X_{1}Y_{2}Y_{3}\cdot X_{1}\cdot Y_{2} \cdot Y_{3}&=1,\\
Y_{1}X_{2}Y_{3}\cdot Y_{1}\cdot X_{2} \cdot Y_{3}&=1,\\
Y_{1}Y_{2}X_{3}\cdot Y_{1}\cdot Y_{2} \cdot X_{3}&=1,\\
X_{1}X_{2}X_{3} \cdot X_{1}Y_{2}Y_{3} \cdot Y_{1}X_{2}Y_{3} \cdot Y_{1}Y_{2}X_{3}&=-1,
\end{aligned}
\end{equation}
where $(\cdot)$ is used to separate operators or operator products. Mermin's argument of the Kochen-Specker theorem is a state-independent proof. In the present time-reversed GHZ experiment, we only identify one ($\ket{\Phi_{0}^{+}}$) out of the eight GHZ states. Thus, for quantum mechanics to interpret the experiment, we have the following eigenequations
\begin{equation}
\begin{aligned}  \label{eigenequations}
X_{1}X_{2}X_{3}\cdot X_{1}\cdot X_{2} \cdot X_{3}\ket{\Phi_{0}^{+}}&=\ket{\Phi_{0}^{+}},\\
X_{1}Y_{2}Y_{3}\cdot X_{1}\cdot Y_{2} \cdot Y_{3}\ket{\Phi_{0}^{+}}&=\ket{\Phi_{0}^{+}},\\
Y_{1}X_{2}Y_{3}\cdot Y_{1}\cdot X_{2} \cdot Y_{3}\ket{\Phi_{0}^{+}}&=\ket{\Phi_{0}^{+}},\\
Y_{1}Y_{2}X_{3}\cdot Y_{1}\cdot Y_{2} \cdot X_{3}\ket{\Phi_{0}^{+}}&=\ket{\Phi_{0}^{+}},\\
X_{1}X_{2}X_{3} \cdot X_{1}Y_{2}Y_{3} \cdot Y_{1}X_{2}Y_{3} \cdot Y_{1}Y_{2}X_{3}\ket{\Phi_{0}^{+}}&=-\ket{\Phi_{0}^{+}}.
\end{aligned}
\end{equation}

How to interpret Eq.~\eqref{eigenequations} by noncontextual  hidden-variable theories? According to Mermin \cite{Mermin:1990:Simple},  each of  operators or operator products (denoted by $O$) separated by $(\cdot)$ can be assigned a predetermined value $v(O)$. Thus the noncontextual hidden-variable theories predict the following relations among these predetermined values:
\begin{equation}
\begin{aligned}  \label{eigenequations1}
v(X_{1}X_{2}X_{3}) v(X_{1}) v(X_{2}) v(X_{3})&=1,\\
v(X_{1}Y_{2}Y_{3}) v(X_{1}) v(Y_{2})  v(Y_{3})&=1,\\
v(Y_{1}X_{2}Y_{3}) v(Y_{1}) v(X_{2}) v(Y_{3})&=1,\\
v(Y_{1}Y_{2}X_{3}) v(Y_{1}) v(Y_{2})  v(X_{3})&=1,\\
v(X_{1}X_{2}X_{3}) v(X_{1}Y_{2}Y_{3}) v(Y_{1}X_{2}Y_{3}) v(Y_{1}Y_{2}X_{3})&=-1.
\end{aligned}
\end{equation}
Since $v(O)=\pm1$, multiplying both sides of Eq.~\eqref{eigenequations1} yields $+1=-1$, which is a conflict. The conflict implies that it is impossible to interpret the experiment by assuming the predetermined values to these operators or operator productions.

There is a trick that the predetermined values of the four operator productions, $v(X_{1}X_{2}X_{3})$, $v(X_{1}Y_{2}Y_{3})$, $v(Y_{1}X_{2}Y_{3})$ and $v(Y_{1}Y_{2}X_{3})$, appear either separately in the first to fourth lines of Eq.~\eqref{eigenequations1}, or jointly in the last line of Eq.~\eqref{eigenequations1}. For the above argument to be valid, either one has to make an additional assumption (e.g., measurements of the four operator productions do not disturb each other) or one has to be able to measure the four operator productions with the same apparatus. A similar argument is essential in a GHZ-like refutation of local realism using two-photon hyperentanglement \cite{Chen:2003:ALL}. Fortunately, in the present case we can avoid the additional assumption also by measuring the four operator productions by the same apparatus, which is exactly the apparatus for the GHZ-state measurement.

The above reasoning is valid for ideal cases, namely, one has $v(O)=\pm1$ exactly and perfect detections. For practical experiments, we have Mermin's inequality \eqref{Mermin's:inequality} by noting that we only identify $\ket{\Phi_{0}^{+}}$ out of the eight GHZ states.

As we noted in Section I.B, in the security proof of our multiparty quantum communication protocols, we suppose that each of Alice, Bob and Charlie has an EPR entangled state which contains one virtual qubit in each of them and the ``signal'' qubit is sent to the middle node, David. After a successful GHZ-state measurement performed by David, the virtual qubit of the legitimate users becomes a GHZ-entangled state. This procedure is known as a multiparty entanglement swapping. If we suppose that each of Alice, Bob and Charlie possesses two EPR-entangled photons, rather than the virtual+signal qubits, a successful GHZ-state measurement by David would result in three-photon GHZ entanglement. The GHZ entanglement created this way can be used to demonstrate the violation of local realism as usual provided that the measurements performed by Alice, Bob, Charlie and David are spacelike separated. Such an experiment can even be performed in a delayed-choice version, as demonstrated for the case of two qubits both theoretically \cite{peres:2000:delayed} and experimentally \cite{ma:2012:experimental}.

\bibliographystyle{apsrev}


\end{document}